\documentclass[useAMS,usenatbib]{mn2e}
\usepackage{graphics,epsf,epsfig,graphicx,longtable}
\usepackage{times}

\newcommand{\AD}{\mbox{ATLAS$^{\rm 3D}$}}

\newcommand{\Reff}{\mbox{R$_{\rm e}$}}

\newcommand{\Mo}{\mbox{M$_{\odot}$}}

\newcommand{\x}{\mbox{$\times$}}

\def\HI{H\,{\small I}}
\def\HI{H\,{\sc i}}

\def\sbr{${\rm mag\,\,arcsec^{-2 }}$\ }

\newcommand\btab[5]{\begin{table*}[#1]{\parbox{#4}{\caption{#2}}\rule[-0.5ex]{0cm}{0.5ex} }
\begin{tabular*}{#4}{#5} }
\newcommand\sbtab[5]{\begin{table}[#1]{\parbox{#4}{\caption{#2}}\rule[-0.5ex]{0cm}{0.5ex} }
\begin{footnotesize}
\begin{tabular*}{#4}{#5} }
\newcommand{\etab}[4]{
\end{tabular*}
\vspace*{#1}
\begin{flushleft}
\parbox{#2}{#3}
\end{flushleft}
\label{#4}
\end{table*} }

\def\rf@jnl#1{{#1}}
\def\aj{\rf@jnl{AJ }}                   
\def\araa{\rf@jnl{ARA\&A }}             
\def\apj{\rf@jnl{ApJ }}                 
\def\apjl{\rf@jnl{ApJ }}                
\def\apjs{\rf@jnl{ApJS }}               
\def\ao{\rf@jnl{Appl.~Opt.}}           
\def\apss{\rf@jnl{Ap\&SS }}             
\def\aap{\rf@jnl{A\&A }}                
\def\aapr{\rf@jnl{A\&A~Rev.}}          
\def\aaps{\rf@jnl{A\&AS }}              
\def\azh{\rf@jnl{AZh }}                 
\def\baas{\rf@jnl{BAAS }}               
\def\jrasc{\rf@jnl{JRASC }}             
\def\memras{\rf@jnl{MmRAS }}            
\def\mnras{\rf@jnl{MNRAS }}             
\def\pra{\rf@jnl{Phys.~Rev.~A}}        
\def\prb{\rf@jnl{Phys.~Rev.~B}}        
\def\prc{\rf@jnl{Phys.~Rev.~C}}        
\def\prd{\rf@jnl{Phys.~Rev.~D}}        
\def\pre{\rf@jnl{Phys.~Rev.~E}}        
\def\prl{\rf@jnl{Phys.~Rev.~Lett.}}    
\def\pasp{\rf@jnl{PASP }}               
\def\pasj{\rf@jnl{PASJ }}               
\def\qjras{\rf@jnl{QJRAS }}             
\def\skytel{\rf@jnl{S\&T }}             
\def\solphys{\rf@jnl{Sol.~Phys.}}      
\def\sovast{\rf@jnl{Soviet~Ast.}}      
\def\ssr{\rf@jnl{Space~Sci.~Rev.}}     
\def\zap{\rf@jnl{ZAp }}                 
\def\nat{\rf@jnl{Nature }}              
\def\iaucirc{\rf@jnl{IAU~Circ.}}       
\def\aplett{\rf@jnl{Astrophys.~Lett.}} 
\def\apspr{\rf@jnl{Astrophys.~Space~Phys.~Res.}}
\def\bain{\rf@jnl{Bull.~Astron.~Inst.~Netherlands}} 
\def\fcp{\rf@jnl{Fund.~Cosmic~Phys.}}  
\def\gca{\rf@jnl{Geochim.~Cosmochim.~Acta}}   
\def\grl{\rf@jnl{Geophys.~Res.~Lett.}} 
\def\jcp{\rf@jnl{J.~Chem.~Phys.}}      
\def\jgr{\rf@jnl{J.~Geophys.~Res.}}    
\def\jqsrt{\rf@jnl{J.~Quant.~Spec.~Radiat.~Transf.}}
\def\memsai{\rf@jnl{Mem.~Soc.~Astron.~Italiana}}
\def\nphysa{\rf@jnl{Nucl.~Phys.~A}}   
\def\physrep{\rf@jnl{Phys.~Rep.}}   
\def\physscr{\rf@jnl{Phys.~Scr}}   
\def\planss{\rf@jnl{Planet.~Space~Sci.}}   
\def\procspie{\rf@jnl{Proc.~SPIE}}   

\title[Early-type galaxies as seen by deep optical images]{The \AD\ project -- XXIX.  The new look of early-type galaxies and surrounding fields disclosed by extremely deep optical images.}

\author
[
Pierre-Alain Duc et al.]{\parbox{\textwidth}{
Pierre-Alain Duc,$^{1}$ \thanks{E-mail:\texttt{paduc@cea.fr}}
Jean-Charles Cuillandre,$^{1,2}$
Emin Karabal,$^{1,3}$
Michele Cappellari,$^{4}$
Katherine Alatalo,$^{5}$ 
Leo Blitz,$^{6}$ 
Fr\'ed\'eric Bournaud,$^{1}$ 
Martin Bureau,$^{4}$ 
Alison F. Crocker,$^{7}$ 
Roger L. Davies,$^{4}$ 
Timothy A. Davis,$^{3}$ 
P. T. de Zeeuw,$^{3,8}$ 
Eric Emsellem,$^{3,9}$ 
Sadegh Khochfar,$^{10}$ 
Davor Krajnovi\'c,$^{11}$ 
Harald Kuntschner,$^{3}$ 
Richard M. McDermid ,$^{12,13}$ 
Leo Michel-Dansac,$^{9}$ 
Raffaella Morganti,$^{14,15}$ 
Thorsten Naab,$^{16}$ 
Tom Oosterloo,$^{14,15}$ 
Sanjaya Paudel,$^{1}$
Marc Sarzi,$^{17}$ 
Nicholas Scott,$^{18}$ 
Paolo Serra,$^{14,19}$ 
Anne-Marie Weijmans,$^{20}$ 
and Lisa M. Young $^{21,22}$}\vspace{0.4cm}\\ 
\parbox{\textwidth}{ 
$^{1}$Laboratoire AIM Paris-Saclay, CEA/Irfu/SAp -- CNRS -- Universit\'e Paris Diderot, 91191 Gif-sur-Yvette Cedex, France\\
$^{2}$Observatoire de Paris, PSL Research University, France\\
$^{3}$European Southern Observatory, Karl-Schwarzschild-Str. 2, 85748 Garching, Germany\\
$^{4}$Sub-Department of Astrophysics, Department of Physics, University of Oxford, Denys Wilkinson Building, Keble Road, Oxford, OX1 3RH, UK\\
$^{5}$Infrared Processing and Analysis Center, California Institute of Technology, Pasadena, California 91125, USA\\ 
$^{6}$Department of Astronomy, Campbell Hall, University of California, Berkeley, CA 94720, USA\\
$^{7}$Ritter Astrophysical Observatory, University of Toledo, Toledo, OH 43606, USA\\
$^{8}$Sterrewacht Leiden, Leiden University, Postbus 9513, 2300 RA Leiden, the Netherlands\\
$^{9}$Universit\'e Lyon 1, Observatoire de Lyon, Centre de Recherche Astrophysique de Lyon and Ecole Normale Sup\'erieure de Lyon, 9 avenue Charles Andr\'e, F-69230 Saint-Genis Laval, France\\
$^{10}$Institute for Astronomy, University of Edinburgh, Royal Observatory, Edinburgh, EH9 3HJ, UK\\
$^{11}$Leibniz-Institut f\"ur Astrophysik Potsdam (AIP), An der Sternwarte 16, D-14482 Potsdam, Germany\\
$^{12}$Department of Physics and Astronomy, Macquarie University, Sydney NSW 2109, Australia\\
$^{13}$Australian Gemini Office, Australian Astronomical Observatory, PO Box 915, Sydney NSW 1670, Australia\\
$^{14}$Netherlands Institute for Radio Astronomy (ASTRON), Postbus 2, 7990 AA Dwingeloo, The Netherlands\\
$^{15}$Kapteyn Astronomical Institute, University of Groningen, Postbus 800, 9700 AV Groningen, The Netherlands\\
$^{16}$Max-Planck-Institut f\"ur Astrophysik, Karl-Schwarzschild-Str. 1, 85741 Garching, Germany\\
$^{17}$Centre for Astrophysics Research, University of Hertfordshire, Hatfield, Herts AL1 9AB, UK\\
$^{18}$Sydney Institute for Astronomy (SIfA ), School of Physics, The University of Sydney, NSW 2006, Australia\\
$^{19}$CSIRO Astronomy and Space Science, Australia Telescope National Facility, PO Box 76, Epping, NSW 1710, Australia\\
$^{20}$School of Physics and Astronomy, University of St Andrews, North Haugh, St Andrews KY16 9SS, UK\\
$^{21}$Physics Department, New Mexico Institute of Mining and Technology, Socorro, NM 87801, USA\\
$^{22}$Academia Sinica Institute of Astronomy \& Astrophysics, PO Box 23-141, Taipei 10617, Taiwan, R.O.C.}}

\begin{document}

\date{Version: final}


\maketitle
\clearpage
\newpage

\begin{abstract}
Galactic archeology based on star counts is instrumental  to reconstruct the past mass assembly of Local Group galaxies. The development of new observing techniques and data-reduction, coupled with the use of sensitive large field of view cameras, now allows us  to pursue this technique in more distant galaxies exploiting their diffuse low surface brightness (LSB)  light. 
As part of the \AD\ project, we have  obtained with the MegaCam camera at the Canada-France Hawaii Telescope  extremely deep, multi--band, images of nearby early-type galaxies. We present here a catalog of  92 galaxies from the \AD\ sample, that are located in low to medium density environments.
The observing strategy and data reduction pipeline, that achieve a gain of several magnitudes in the limiting surface brightness with respect to classical imaging surveys,  are presented. The size and depth of the survey is compared to other recent deep imaging projects.
The paper highlights the  capability of LSB--optimized surveys at detecting new prominent structures that change the apparent morphology of  galaxies.
The intrinsic limitations of  deep imaging observations are also discussed, among those, the contamination of the stellar halos of galaxies by extended ghost  reflections, and the cirrus emission from  Galactic dust.
The detection and systematic census  of fine structures   that trace the present and past mass assembly of ETGs is one of the prime goals of the project. We provide  specific examples of each type of observed structures  -- tidal tails, stellar streams and shells --, and explain how they were identified and classified. 
We give an overview of the initial results. The detailed  statistical analysis will be presented in future papers. 

 \end{abstract}


\section{Introduction}
\label{sec:intro}
Our knowledge of nearby, well resolved, galaxies made a big leap during the last 40 years with the availability of multi wavelength data, whereas for the previous decades only optical data, mostly images, were at the astronomers' disposal. Nowadays a nearby galaxy can no longer be characterized without ultraviolet and far infrared  data revealing its  star--forming activity, a near infrared  image constraining its stellar mass, radio and millimeter maps, providing information about its  gas content. Consequently, the interest for pure optical imaging surveys of galaxies has dropped, unless such surveys cover large regions of the sky and provide statistical information -- like the Sloan Digital Sky Survey \citep[SDSS,][]{York00} --, reach  high angular resolution and give insight into  nuclear regions, or target the high redshift Universe which still largely lacks  non-optical data.

However, several recent  imaging surveys benefiting  from innovative observing techniques and instruments have rejuvenated our regard for the optical regime and to familiar galaxies, disclosing around them so far unknown  prominent but low surface brightness (LSB) stellar structures,  such as   extended stellar halos and tidal tails \citep[e.g.][]{Mihos05,Janowiecki10,Martinez-Delgado10,Roediger11,vanDokkum14}.
At the same time, the development of cosmological numerical simulations contributed to foster the interest for deep imaging of the nearby Universe.
Following the paradigm of the hierarchical model,    they figure out that todays massive galaxies grew from  series of galactic collisions that left  around them various types of vestiges, including shells and tails \cite[e.g][]{Bullock05,Naab07,Oser10,Helmi11,Cooper14}. The technique known as galactic archaeology  that makes a census of LSB collisional debris with star counts has so far mostly been  applied to galaxies in the Local and very nearby groups
\cite[e.g][]{McConnachie09,Crnojevic13}.  Further away, structures can no longer be resolved into stars and other  excavation tools should be used, in particular the diffuse  light. 

Whereas the large field of view photographic plates used in the 1950s-1970s  enhanced with amplified techniques \citep{Malin78}  were capable of detecting the extended LSB component of galaxies,  the  early CCDs   developed in the 1980s, with their   limited field of view, proved to be much less efficient at that task.  
The  complex cameras made with multiple optical elements that were built to host these new sensitive detectors, while well fitted to detect distant galaxies, generate numerous artefacts, such as internal reflections that hide extended LSB structures. 
Thus somehow ironically LSB science made a step backwards with the advent of CCDs, and the few extragalactic papers  published on the topic during the period 1980--2000 mostly focussed on the outer stellar populations of spiral galaxies  \cite[e.g.][]{Lequeux96}. 
 Only recently    astronomers, among them amateurs  observing in very dark sites  with simple cameras,  raised the challenge of detecting again the diffuse light \citep{Martinez-Delgado09}. The astonishing images they produced pushed professional astronomers to develop new  techniques to eliminate the instrumental signature in their camera. This involves special coating of the detectors \citep{Mihos05},  LSB-optimized  observing techniques with  large field of view mosaic cameras \citep{Ferrarese12}, or even the construction of new LSB dedicated cameras \citep{vanDokkum14}.

The gain of deep imaging is for some galaxies tremendous, with the detection of  networks of interlaced filaments \citep{Martinez-Delgado10,Paudel13}, revealing past  mass accretion  histories that were much more complex than initially thought. However the sample of nearby objects with available deep optical imaging remains limited and highly biased towards galaxies for which previous imaging surveys such as the SDSS already indicated the possible presence of collisional debris \citep{Miskolczi11}. 
This was the motivation for carrying out  a systematic deep imaging survey of a well selected  large sample of galaxies, allowing us to determine how the properties of the outskirts of galaxies vary  with morphology, colour, gas content and structural parameters such as the mass, size and dynamics. 
A  survey of several tens of objects implies observing with at least medium size telescopes rather than small amateur-type telescopes, to keep individual exposure times (and thus survey length) relatively small, typically one hour instead of a full night. Using professional facilities ensures as well good photometric accuracy and image quality. 

As part of the  \AD\ project \citep{Cappellari11}, we have carried out with the Canada-France-Hawaii Telescope (CFHT) and the  MegaCam camera a deep multi-band imaging survey of  nearby Early-Type galaxies (ETGs).
This paper presents a  comprehensive  image catalog of 92 objects, located in  low to medium density environments. Among them,  a few systems were devoted individual studies \citep{Michel-Dansac10b,Duc11,Serra13,Alatalo14,Duc14}.

We  address here the global survey strategy, observing and data reduction technique, and discuss its  limitation, including contamination by instrumental artefacts such as diffuse halos of  bright stars  and galactic nuclei, as well as by foreground Galactic cirrus. These effects are not necessarily specific to images obtained with MegaCam, and may affect any deep imaging survey. 
The gain of several magnitudes in limiting surface brightness compared to previous generation of imaging surveys makes us enter into a new regime having specific stumbling blocks which cannot be ignored and should be investigated in detail. 

Sect.~\ref{sec:obs} presents the observations at CFHT,  the observing strategy and data reduction technique. Sect.~\ref{sec:perf}  discusses the survey performance and limitations.  Sect.~\ref{sec:prod}  details the sample selection as well as the high level image production. 
The image catalog  consists of  composite multi-band images with a true colour rendering, surface brightness plus colour maps,  and residual images, i.e. images in which the galaxy has been modeled and subtracted.
Sect.~\ref{sec:class} discusses the need to (re-) classify galaxies based on the newly discovered structures: extended halos, discs, stellar streams and tails, shells.
Conclusions and perspectives are given in  Sect.~\ref{sec:sum}. 
The scientific analysis of the images  and statistical results will be presented in future papers.

\section{Observations and data reduction}
\label{sec:obs}

Observations have been carried out at the Canada-France-Hawaii Telescope, with the MegaCam camera,  as part of a series of  regular multi--semester PI  and snapshot programs in the framework of the \AD\ project (10BF11, 11AF06, 11AD89, 11BD93, 12AF04, 12AF99). 
The galaxies presented in this  image release  were observed between 2010 and 2013.

\subsection{Context}
MegaCam \citep{Boulade03} is a wide-field imager, made of 36  CCDs and covering 1 square degree field of view.  It was initially designed for point-source type science, stars and distant galaxies, where large scale structures of the image background are secondary to photometric accuracy across the field of view. As part of the CFHT Legacy Survey effort, refinements over the years led to a better than 1\% photometric accuracy across the one degree field-of-view, a precision of great importance when dealing with precision cosmology \citep{Regnault09,Betoule13}. All these advances were integrated in the CFHT MegaCam official pipeline \citep{Magnier04} over the years. Since the background is usually internally subtracted for point-source type science, the global gradient caused by diffuse sky background reflections in the optics, plus the cumulative radial effect of the illumination correction to reach the percent photometric precision over the entire field of view, is of little concern.

But since the instrument first light in 2003, efforts were made to also enable scientific programs dealing with large and faint extended components that are de facto lost within the background of the image. The overall effort aimed at recovering the true sky background, that is a purely flat response  on top of the astronomical sources signal.
The nod-and-subtract technique used in near-infrared astronomy was an inspiration for the observing strategy and pipeline developed for MegaCam, called Elixir-LSB. In the near--infrared regime, though, the background dominates the signal and varies over relatively short time scales.  In the optical the idea is to model a background stable in time and dominated with astronomical sources, with just a handful of exposures. 

We note that the processing presented in the following does not alter the photometric accuracy at small scales, i.e. the 1\% photometry precision is retained throughout the background correction process.

\subsection{Dithering, sky stability and overheads}
\label{sec:strategy}

In the g-band and the r-band, studies have demonstrated  that the sky background on Mauna Kea is stable at a fraction of a percent over timescales of 1 hour, as long as the Moon does not rise or twilights are close in time. Since we want to model a background map common to, and built from, a series of consecutive images, we must limit the observing sequence in time. It is also important to have a sky photon limited regime in each exposure and reach a reasonable balance with the observing overheads (the camera readout time is 40 seconds). This led us to a single exposure integration time of $\sim$5\,mn. To reach 29 magnitudes per square arcsecond with Elixir-LSB, early studies demonstrated that $\sim$40\,mn total integration is enough (beyond that point, the systematics dominate - see below for a discussion).

However, the Elixir-LSB technique requires at least 7 images to derive a proper map of the background, with a telescope offset between exposures large enough to skip over the largest features caused by the astronomical sources: extended galaxies, and reflection halos from bright stars (7 arcmin). At the same time we want the main target of interest to be integrated 100\% of the time: luckily with the large field of view of MegaCam and the apparent size of the galaxies in the \AD\ sample,  all  features are smaller than 10 arcmin, and  can thus be moved within an extended central region of the mosaic while enabling the background subtraction process. This is different from  the technique used for the Next Generation Virgo Cluster Survey (NGVS) where the need to map large areas led to a different observing strategy  \citep{Cuillandre11,Ferrarese12} 

\begin{figure}
\includegraphics[width=\columnwidth]{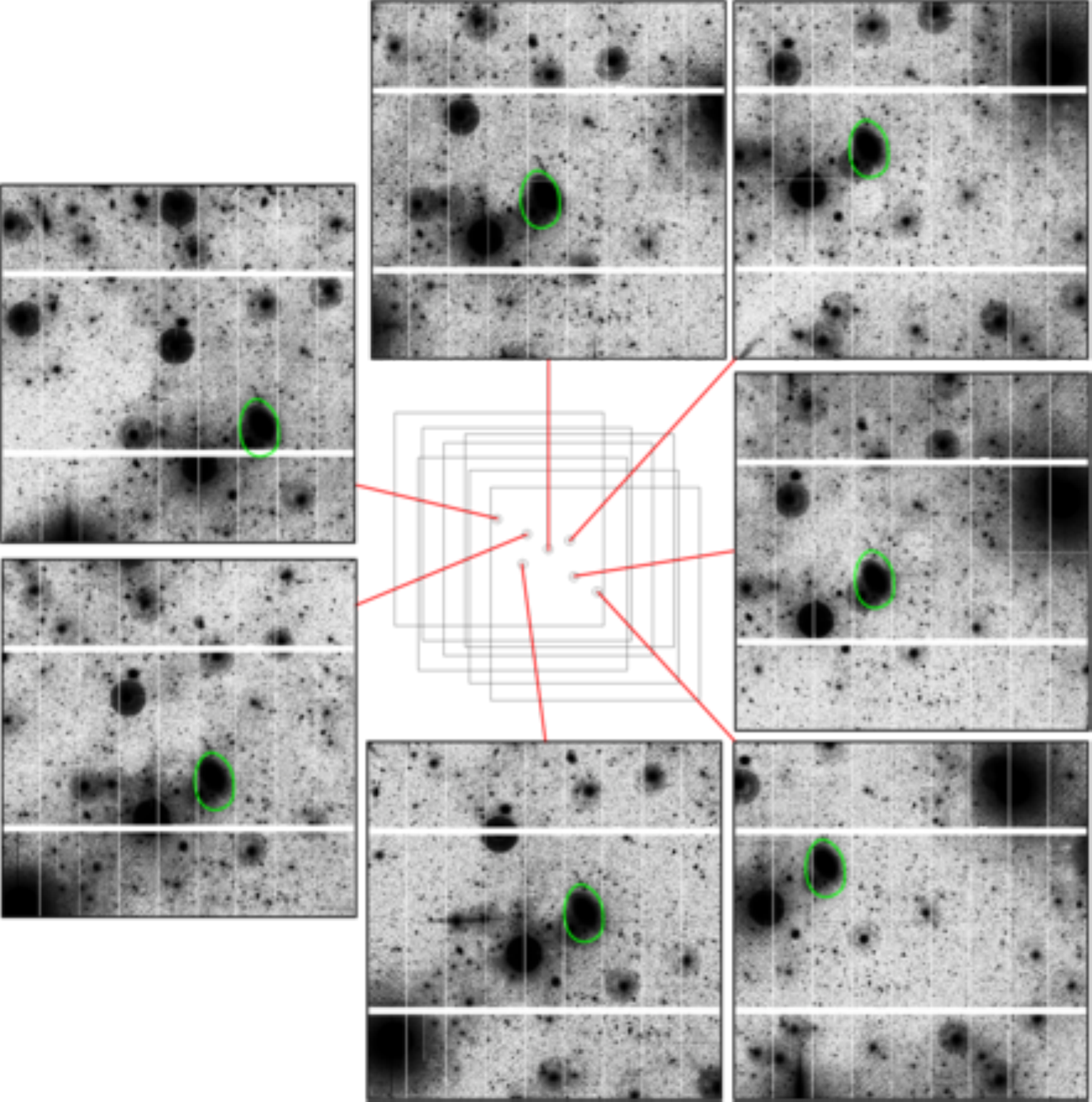}
\caption{The adopted observing sequence.  The offset pattern is shown on the middle panel.  The  adjacent panels display the 7 individual g--band sky-subtracted images  before  stacking.  
The target galaxy, here NGC~5582, is surrounded by the green ellipse.}
\label{fig:obs-seq}
\end{figure}

A specific dithering pattern was implemented in the CFHT service observing system to serve our survey, but it also applies to many other programs with sources of similar physical scales. That pattern with offsets in RA and DEC ranging between 2 and 14 arcmin allows the galaxy to never occupy the same physical area of the CCD mosaic across the 7 exposures, and in consequence allows the construction of a background map.
 The pattern is also designed such that the mosaic gaps are naturally removed when stacking the images. The  offset pattern is shown  in Fig.~\ref{fig:obs-seq}. An example of the observing sequence is illustrated in the same figure.

\begin{figure}
\includegraphics[width=\columnwidth]{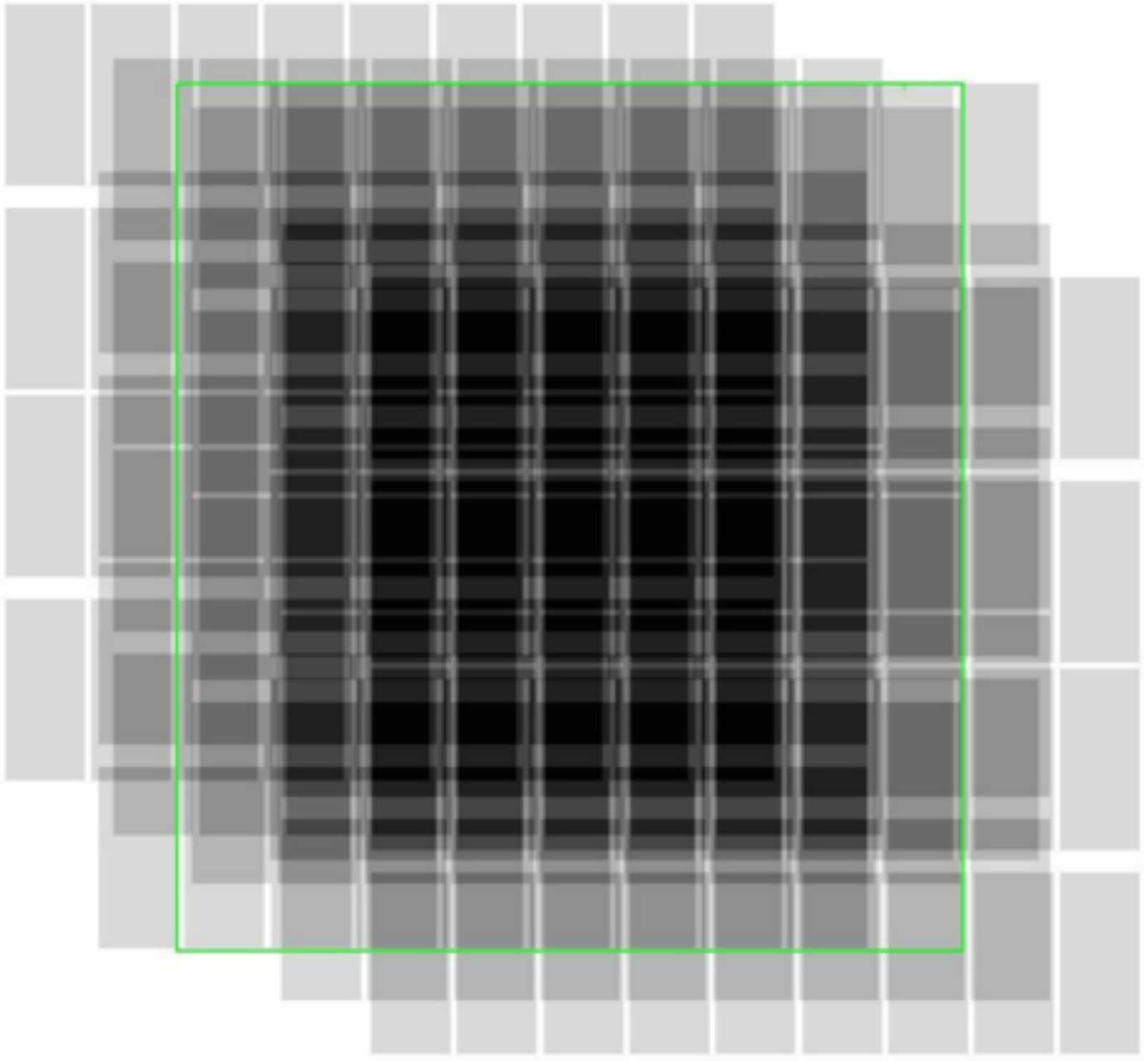}
\caption{Weight map resulting from  the stacking process. Darker regions have a higher pixel redundancy and thus sensitivity. The green square corresponds to the area kept for the final stacked image (63 $\times$ 69 arcmin).}
\label{fig:obs-weight}
\end{figure}

Each MegaCam pointing covers one square degree, and the field resulting from the  image recombination with our LSB-optimized technique would yield  a much larger one, but we limit our stack sky coverage to the central 63 $\times$ 69 arcmin where  the  signal-to-noise is the highest, though still with a lower sensitivity at the edges. The resulting weight map is shown in Fig.~\ref{fig:obs-weight}.

On the stacked image, the target galaxy is located close to the centre of the field, precisely at position +0.1',+2.1' with respect to the centre. Within the target field of view, other \AD\ galaxies, most often lying in the same group as the primary, may be present. In that case, observations were generally not duplicated. Depending on their location, these secondary targets have images with a slightly lower sensitivity (but see below); the precise location of each object within the original  frame is shown in the on--line version of the catalog.  Due to technical issues with the telescope scheduling system, a few galaxies were observed twice and their stacked images were made with 12--14 individual exposures instead of 7.  Conversely, observations made  during the first observing run consisted of only 6 individual exposures. The number was adjusted to 7 for the following runs to optimize   the background subtraction. 

\subsection{Background correction}

The basic idea is to derive a map of the background which is mostly a large radial pattern caused by the sky background reflection in the optics and the photometric illumination correction to deliver a photometric flatness across the field of view. The entry frames for the Elixir-LSB pipeline are the regular Elixir frames where that radial structure is left untouched. The amplitude of that radial gradient is nearly 15\% of the sky background.

A multitude of astronomical sources of different angular sizes are present in the background. With the pattern selected we ensure that a column of pixels across the 7 images at any given location on the CCD mosaic will mostly see the sky (no astronomical sources). Extensive testing has demonstrated that this is true for any sky area with a high enough galactic latitude.

The 7 frames are median-stacked and the resulting image is smoothed on a 4 arcmin scale to reject possible artefacts due to a higher number of astronomical sources in the column of images at a given location. This map is then scaled back to each image's sky background level, and subtracted. The result is an image  with sky background perfectly flat, with residuals left from overcrowding over an area of the CCD mosaic across the 7 images. In that case, there will be a slight overcorrection of the background, leaving a  dip in the corrected image. However in the worst case, the deviation is not more than $\sim$0.4\% of the sky background (max-min/background). We thus consider that 7 exposures achieve a  satisfactory level. 

\begin{figure}
\includegraphics[width=\columnwidth]{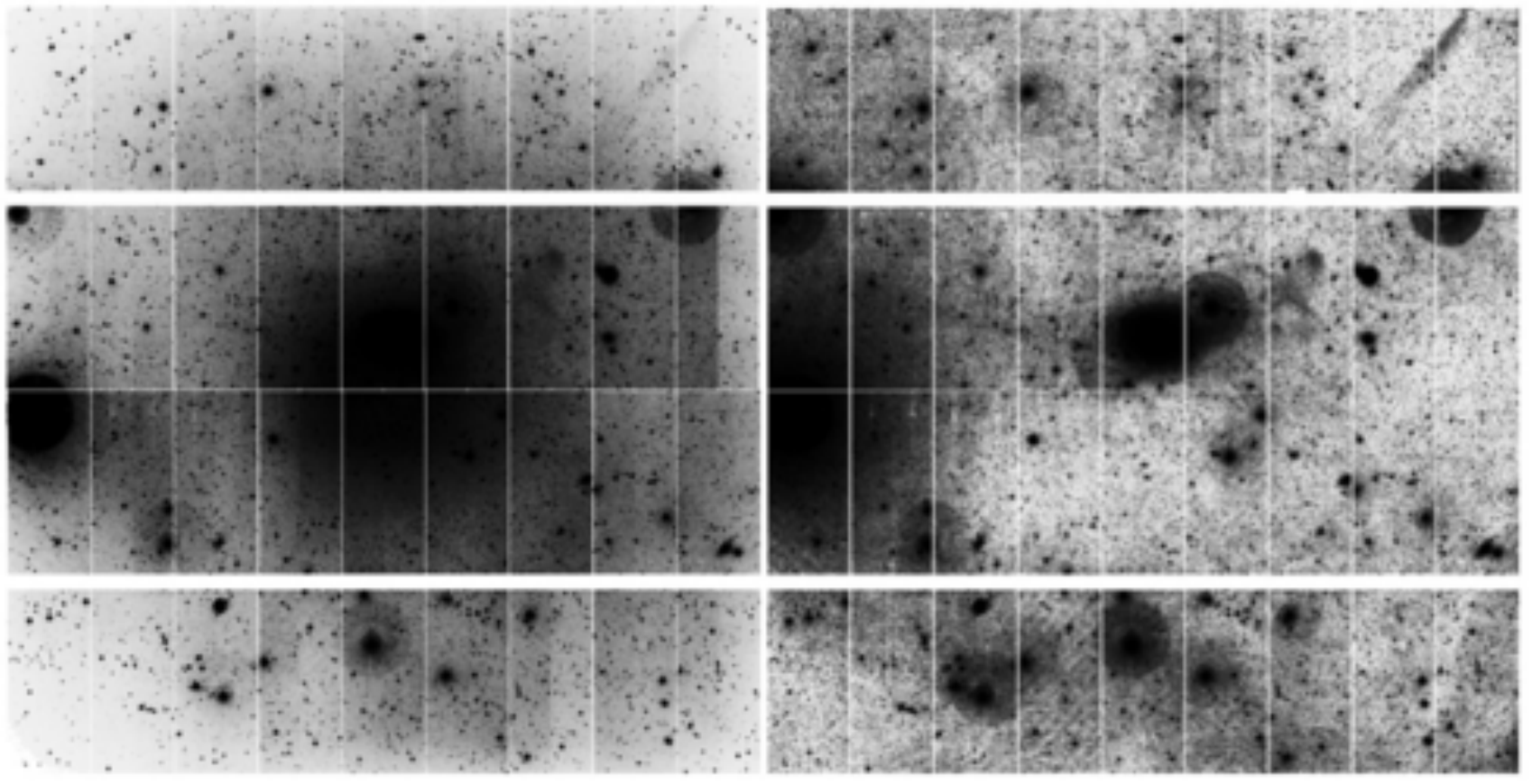}
\caption{The dramatic gain of the LSB-optimized observing strategy and data reduction. The left panel shows one individual g--band exposure of  NGC~5557 for which a bias and standard flat-field correction has been applied. At the chosen intensity contrast, the central regions of the frame are polluted by an extended scattered emission. On the  individual image processed by Elixir-LSB  shown on the right panel, the central scattered light has disappeared and the faint LSB tidal features around the galaxy show up. The CCD patterns which are  visible on this figure are removed by the final image stacking.}
\label{fig:skysub}
 \end{figure}

At the end of the background correction process, each of the 7 images now has a background flat within 0.4\% versus the 15\% initially. This means the true sky background (flat, at least in g and r) has been restored. An example of a background subtracted individual image, illustrating the  efficiency of the method,  is shown in Fig.~\ref{fig:skysub}.

\subsection{Image stacking}

The 7 images are run through the AstrOmatic astrometry package {\sc SCAMP} \citep{Bertin06}, along the other set of images of the given target to ensure a uniform solution. The {\sc SCAMP}  output is used by the AstrOmatic resampling package {\sc SWARP}  \citep{Bertin10} with the image sampling going from 0.19 arcsec per pixel to 0.54 in order to facilitate the processing speed, but mostly to boost the signal-to-noise ratio with the filtering effect of the 3 by 3 average binning. The overall astrometry precision is about 0.05 arcsec (1/10th of a pixel).

The sky background (now the true sky with a constant value across the image) is precisely measured and subtracted. The final step consists in stacking the frames using a standard signal-clipping approach to reject artefacts (satellite tracks, cosmic rays).
Stacking the frames by averaging areas of the mosaic which have seen different parts of the sky further decreases the artefacts left by "vertical" astronomical crowding in the column of images when building the background. Elixir-LSB stacks deliver for MegaCam images flat at the $\sim$0.2\% level of the sky background.

The absolute photometric calibration produced by Elixir is tracked through the entire process and the stacks are calibrated in the MegaCam AB natural magnitude system from where transformation to other systems such a Sloan are readily available.

\subsection{Limitations of the approach}

0.2\% of the sky background means it is possible to detect low surface brightness features some 7 magnitudes fainter than the sky background. That limit has proven to be the same across all 5 broad band filters available on MegaCam. For the g-band, this corresponds to a 28.5 magnitude per square arcsecond limit.

Because that limit of 0.2\% is the same across all bands, this hints that the approach is limited by systematic errors at that level. Indeed, deeper integration provides only minor gains in flatness. The origin of the problem lies in the optical nature of the instrument with a 4-lens wide-field corrector added to the regular optics of the camera, leading to a large number of possibilities of internal light reflection. Here the problem is not the sky background but the many sources seen in the camera beam that all create their own set of halos, most of them too faint to be seen as the ones caused by bright stars, but all contributing to making the background uneven enough between exposures. Effects of these halos on the science results are discussed in Sect.~\ref{sec:halos}.

There is an on-going effort in modeling the halos in MegaCam for each individual exposure and correct them to further boost the performance of Elixir-LSB (Cuillandre et al., in preparation). If that effort is conclusive surely the whole survey would gain in depth, possibly reaching the $\sim$30~\sbr limit.

\subsection{Filter selection and observing time}

The choice of filters   was primarily driven by our aim to detect low--surface brightness features and extend the study of the stellar populations of ETGs to large radii (up to 10~\Reff). We have selected the  filters that maximize the contrast between the  sky background and the debris while keeping the exposure time reasonable: g' and r' \footnote{The u*, g', r', i', z' filters of MegaCam slightly differ from the Sloan u, g, r, i and z filters, hence their different naming. However, for simplicity, throughout the paper, we will refer to them as u, g, r and i.}.  From the derived g-r colour map, constraints on the stellar age and metallicity gradients may  be obtained, though with a large degeneracy. 
As argued in Sect.~\ref{sec:FS}, colour information is also useful to distinguish  tidal tails formed in major mergers, composed of mixed, metal--rich,  stellar populations expelled from their parent spiral galaxies, to stellar streams that result from the disruption of low--mass satellites primary composed of low--metallicity stars.

In addition, i band images were obtained for a sub-sample of ETGs, allowing in particular a comparison of the colour profile with galaxies located in the Virgo Cluster which were also observed with MegaCam as part of the NGVS  but for which the r band is missing.
Finally, u band images were acquired for a few  ETGs with distances  below 20~Mpc.  The u band helps to identify  star--forming regions in particular in gas--rich outer regions of ETGs such as discs and tidal debris. However, the primary motivation for the u band observations is the detection of globular clusters (GCs). Combining u with  g, r and i  helps to identify GCs against foreground stars and background distant galaxies (Lan\c{c}on et al., in prep). 
These multi-band images were combined to compute the ``true colour" images shown in the catalog. 

Given the observing strategy defined in Sect.~\ref{sec:strategy}, exposure times were  typically 7 $\times$ 345 sec in g and r bands,  7 $\times$ 230 sec in  the i band and 7 $\times$  700 sec in the u band. The list of available bands, total exposure times, number of individual exposures, image quality and background level for the first galaxies in the catalog is given in Table~\ref{tab:obsmeg}. The full table with all observed galaxies is available in the online web version.

\begin{table}
\caption{Catalog of ETGs with MegaCam observations (first entries; full table available in the online web version)} 
\begin{tabular}{lccccl} 
\hline 
Galaxy & Band & N & Integration time & IQ & Background \\ 
& & & s & arcsec & ADUs \\ 
(1) & (2) & (3) & (4) & (5) & (6) \\ \hline \\ 
\hline 
NGC0448 & g' & 7 & 2415 & 0.64 & 704.29 \\ 
& r' & 7 & 2415 & 0.77 & 941.57 \\ 
& i' & 7 & 1610 & 0.65 & 1054.86 \\ 
\hline 
NGC0474 & u* & 7 & 4900 & 1.16 & 223.43 \\ 
& g' & 7 & 2415 & 0.83 & 784.43 \\ 
& r' & 7 & 2415 & 0.66 & 935.43 \\ 
& i' & 14 & 3220 & 0.65 & 1080.79 \\ 
\hline 
NGC0502 & g' & 7 & 2415 & 0.82 & 688 \\ 
& r' & 7 & 2415 & 0.79 & 981.29 \\ 
& i' & 7 & 1610 & 0.66 & 1626.29 \\ 
\hline 
\multicolumn{6}{l}{\parbox{8cm}{Notes: (3) Number of individual exposures (4) Total integration time (5) Image Quality: FWHM of the PSF (6) Background level}} \\ 

\end{tabular}
\label{tab:obsmeg} \\
\end{table}

\subsection{Observing conditions and image quality}

All observations were obtained in dark time, far from the twilights, so as to maximize the signal-to-noise of LSB structures and avoid varying sky background. 
A large fraction of our images were obtained with seeing better than 1 arcsec for u, g, r and 0.7 arcsec for i, i.e. with observing conditions that are not needed for LSB science.  The Image Quality of the MegaCam survey enables to address   additional scientific objectives, such as  the identification of  Globular Clusters around the ETGs or the estimate of their distance with the technique of surface brightness fluctuations. 
Thus one of the main values of the MegaCam deep imaging project is to  benefit from an observing strategy optimized for the detection of LSB structures while  also keeping the  excellent spatial resolution of MegaCam and seeing of the Mauna Kea.

\section{Deep imaging survey performances and limitations}
\label{sec:perf}

We discuss here the  performances and limitations of our  MegaCam survey. Note that some of the issues raised here, in particular the extended reflection halos around bright objects and Galactic cirrus emission, are of broad interest as they plague similar  deep imaging surveys.   

\subsection{Instrumental signatures} 

Some  instrumental signatures remain visible after the Elixir-LSB and stacking processes.  They are invisible when cutting the images at 27 \sbr, but become prominent at fainter surface brightness limits. 
We discuss here how they affect the science analysis and propose ways to minimize them. 

\begin{figure}
\includegraphics[width=\columnwidth]{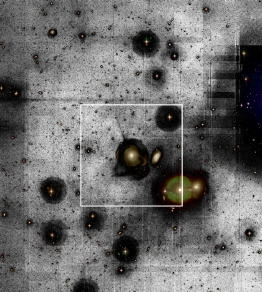}
\caption{g--band image of the field around  NGC~474. The chosen intensity scaling enhances the fainter extended low surface brightness features, including the extended reflection halos around the bright stars and other instrumental signatures, such as horizontal and vertical bands corresponding to lower sensitivity in the CCD gaps and at the edges of the recombined frames. On such images, non circular or rectangular features, and oblique filaments  are real structures on the sky. The target galaxy lies at the centre of the white square, the size of which corresponds to 40 times the effective radius of the ETG.  To better identify the objects, a composite g+r+i   image has been superimposed.
The field of view of the large-scale image is 57 \x 64 arcmin. North is  up and East left. 
 Similar images are provided for all our targets in the image catalog.}
\label{fig:inst}
\end{figure}

\subsubsection{CCD gaps}
MegaCam is a mosaic of 36 CCDs with  gaps of 13 to 80 arcsec between them. On the stacked images, their signature are horizontal and vertical bands of higher noise (see examples in Fig~\ref{fig:inst}). Such geometrical patterns are easily identified and most often cannot be misidentified with real celestial objects. Besides, they do not affect much the photometric measurements.

\subsubsection{Stellar halos} 

\label{sec:halos}

\begin{figure}
\includegraphics[width=\columnwidth]{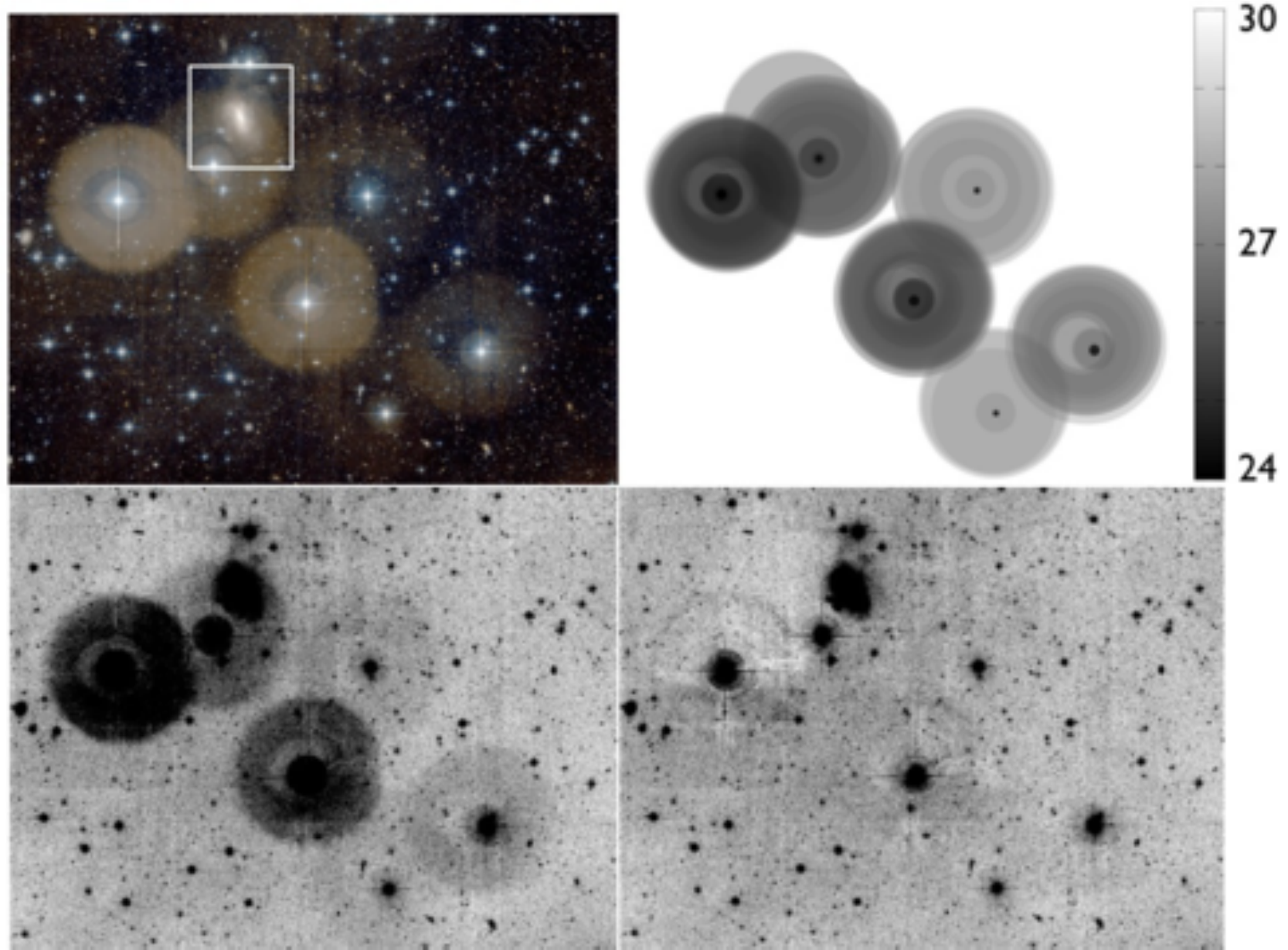}
\caption{Imprints of the internal reflections of bright stars on deep images with MegaCam.  {\it Top-left:}  composite g+r--band image of the field around NGC~2764. The position of the ETG is indicated by the white square.  {\it Top-right:} empirical model of the reflections in the r--band. A surface brightness scale, ranging between 24 and 30 \sbr, is used.  {\it Bottom-left:}  original r--band image. {\it Bottom-right:} resulting  image at the same intensity scale after subtraction of the stellar halos.}
\label{fig:star-halos}
\end{figure}

\begin{figure}
\includegraphics[width=\columnwidth]{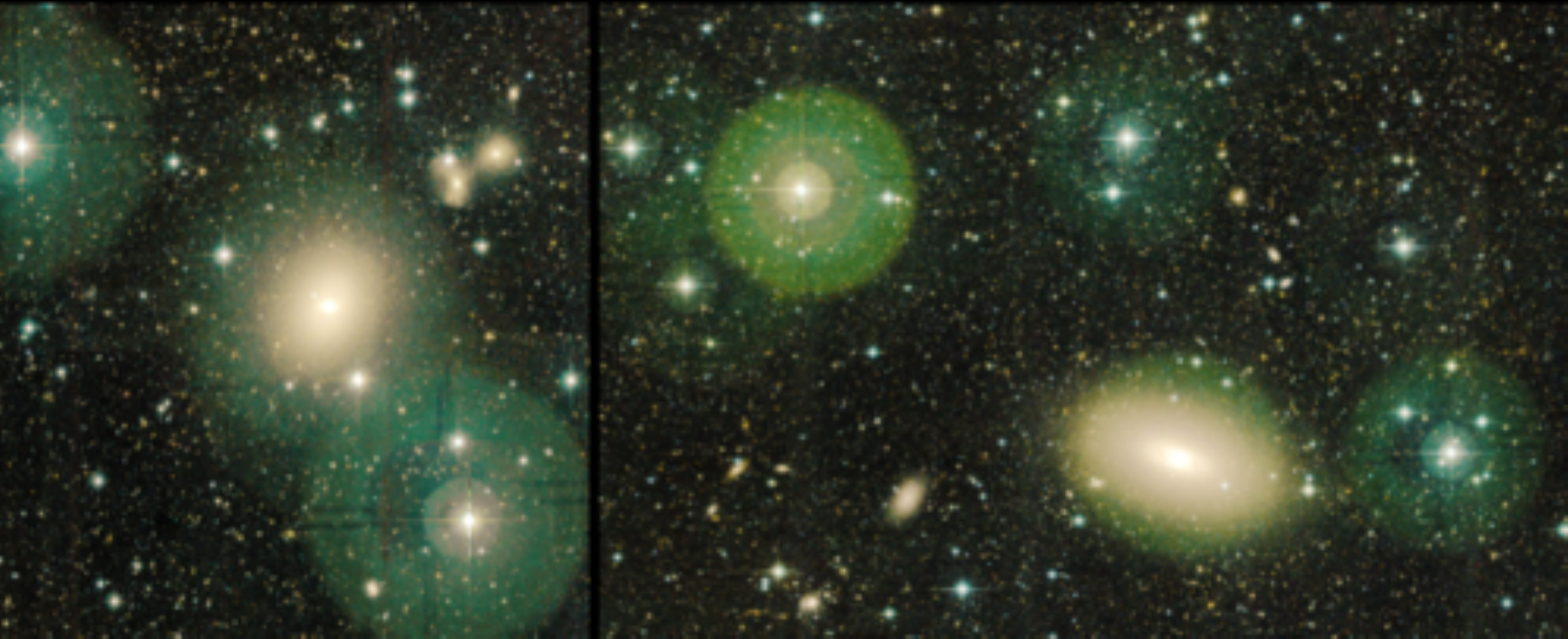}
\caption{Extended low surface brightness nuclear and stellar halos as seen on composite g+r+i images of  the galaxies NGC~5473 (left) and NGC~3489 (right).
 The nucleus of each galaxy generates its own ghost halo which has a size and colour roughly similar to that of the surrounding bright stars.}
\label{fig:halos}
\end{figure}

More detrimental are the  halos of bright stars, as their circular shape gets closer to that of real galaxies, especially those seen face-on. 
They are superimposed on the  classical wings of the instrument  PSF.
The outer envelope of these  features resembling out of focus stellar images  has a typical  radius of 3.5 arcmin. 
The halos are of low surface brightness: most of them have mean surface brightness in the g band fainter than 26.5~\sbr (see Fig.~\ref{fig:star-halos}). 
However, in LSB optimized deep  images, they occupy large areas. As an example, in the MegaCam  image shown in Fig.~\ref{fig:inst}, visible halos cover about one fifth of the frame. As a consequence these extended structures often spatially overlap with the outer regions of galaxies and contaminate their photometry.
Such halos, present in many imaging instruments, are imprints of  internal reflections between the CCDs and different optical elements of the camera, including the dewar window and filters  \citep{Slater09}. 
They have complex shapes: the multiple internal reflections produce several more or less concentric overlapping discs; the exact position of the central star with respect to them depends on its position on the frame. 
Furthermore, their brightness depends on the filter used; in our survey they are mostly prominent in the r--band  and appear in green on our composite g+r+i images (see Fig~\ref{fig:halos}).
Indeed, the reflectivity of MegaCam has been minimized in the blue domain. At longer wavelength (i--band and beyond), the halos are lost in the  sky background, so that their brightness peaks in the r--band.

Subtracting these halos is  a tedious but necessary task. 
We have modeled and removed the most prominent stellar halos located close to the target in an empirical and interactive way. We fitted the external reflection pattern with a disc of constant brightness, which we subtracted from the image. We iterated  fitting the residual image with another disc of smaller size and slightly different centre. The process was pursued until only the central brightest - saturated - spot remained on the residual. 
This halo subtraction requiring manual interaction was particularly time consuming, but proved efficient in revealing the structures of galaxies even rather close to bright stars. It basically leaves only the imprint of the  obstructing  support spider of the telescope mirror (see Fig.~\ref{fig:star-halos}). 

As mentioned earlier, the exact shape of the halos depends on the position of the star in the MegaCam field. Our final image is a combination of individual frames  which were shifted on the sky by large offsets. This resulted in stellar halos with an even more complex structure than on an individual frame. Ideally, the halos should be removed in  the 7 individual  frames before stacking, multiplying by the same amount the time necessary to remove them. To avoid  this, the stellar halos were computed from the final image and models were slightly blurred.

\begin{figure}
\includegraphics[width=\columnwidth]{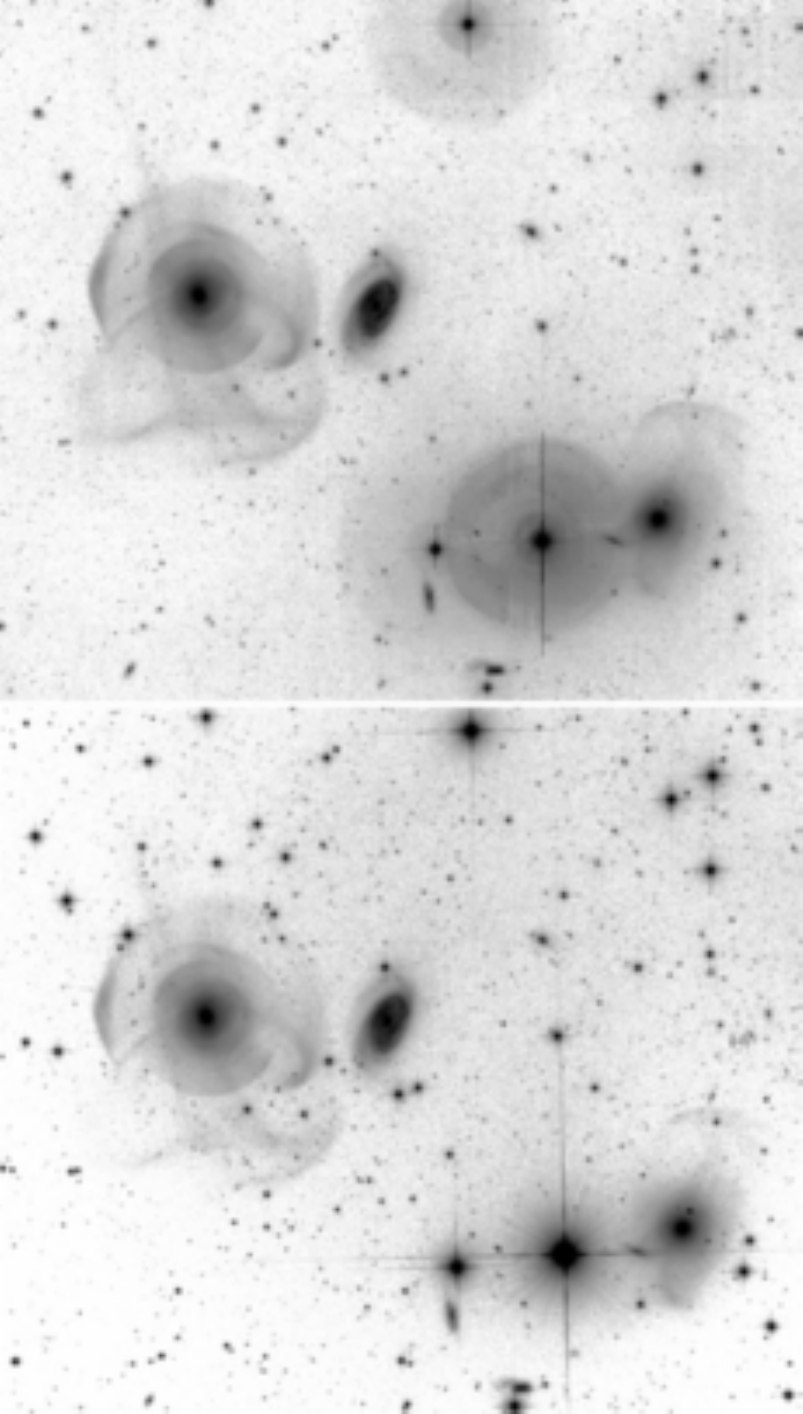}
\caption{Images of the same field around NGC~474 (to the East) and NGC~467 (to the West) obtained with two different cameras and telescopes:  MegaCam on the 3.6m CFHT (top) -- total exposure of 0.7 hour  in the g-band -- and an ATIK 4000m CCD camera mounted on a 12" RC amateur telescope, located on the site of the  Bulgarian National Astronomical Observatory Rohzen  -- total exposure time of 21.5 hour with the Clear Luminance filter (L);  Image credit:  Irida telescope, Velimir Popov and Emil Ivanov -- Note the absence of  large stellar halos in the image obtained by the Irida telescope. 
}
\label{fig:pro-amateur}
\end{figure}

Some surveys have circumvented this difficulty by minimizing light scattering and reflections with light-absorbing material   and antireflective coatings inside the telescope and the camera \citep{Mihos05,Slater09}, or developing  concept cameras with radical different technologies \citep[see the Dragonfly Telephoto Array,][]{vanDokkum14}. Images generated by amateur cameras with simple optics also suffer much less from internal reflections and produce images that at first sight may be considered to be cleaner. An instructive comparison between the CFHT image of the field around the galaxies NGC~474 and NGC~467 obtained with CFHT/MegaCam and with a small 12 inch (0.3~m) telescope as part of a  collaboration with amateur astronomers is presented in Fig.~\ref{fig:pro-amateur}. Most of the LSB features revealed by the MegaCam camera are visible on the amateur image, though the sky on the amateur image is not perfectly flat. Obviously, compared to cameras installed on amateur telescopes or just made of  telephoto lenses,  MegaCam on the CFHT benefits from the use of a 4-meter class telescope, allowing to reach a similar surface brightness limit in 30 times less observing time.

\subsubsection{Nuclear halos} 
A potentially more devastating effect of internal reflections,  which  until recently has  been often minimized, is the radial spreading of the brightest parts of galaxies, i.e. their nucleus, at large effective radius. This PSF far wing generates structures which may be mistaken for real stellar halos \citep{Michard02,Sandin14}. With shallow imaging surveys, the PSF signature may be seen in stacked images 
\citep{deJong08,LaBarbera12,Dsouza14}. With our deep imaging survey, it is directly visible on individual images. 
Fig.~\ref{fig:halos} displays composite g+r+i images of the fields surrounding NGC~5473 and NGC~3489.  It is remarkable that the outer structures of these galaxies share the same roundish shape, size and even colour as the reflection halos due to the nearby bright stars.  They have in fact the same origin. The compact  galactic nuclei generate  halos which are just a little more fuzzier than those  produced by point--like stars.

\begin{figure}
\centerline{\includegraphics[width=0.8\columnwidth]{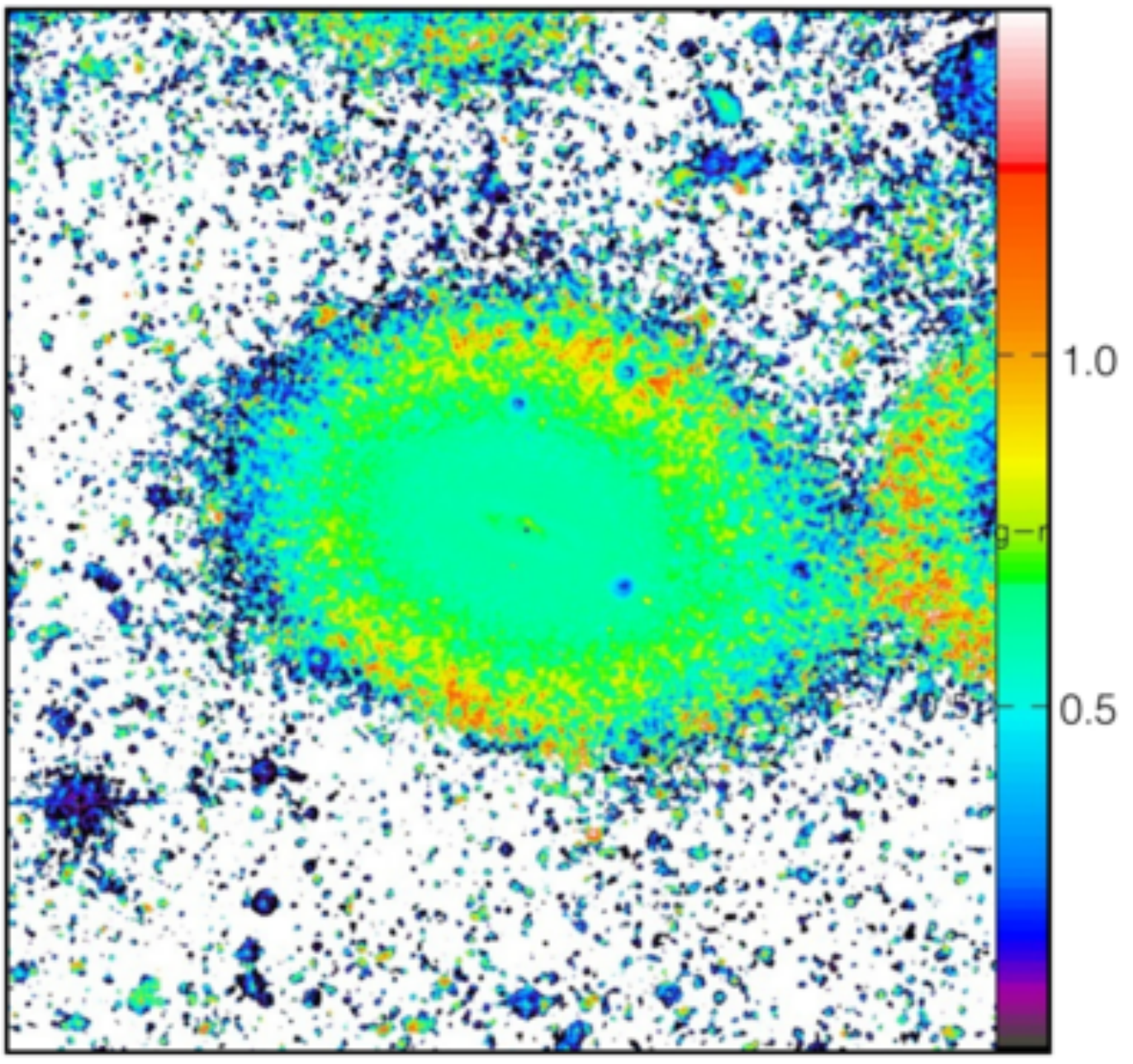}}
\caption{ g-r colour map of the galaxy NGC~3489. The reddening by 0.5~mag in the outer  annulus is most likely  due to the internal reflection of the red galaxy nucleus. The scale in mag is shown to the right.}
\label{fig:halo-color}
\end{figure}

In the case of NGC~5473, the fake reflection halo is easily identified as it is larger than the stellar halo of the galaxy and has a different almost monochromatic (green)  colour.  For  NGC~3489, the ghost and stellar halos have about the same size, but because of  the large ellipticity of the galaxy, the  fake halo is still visible  along the minor axis. When the apparent size of the galaxy exceeds that of the reflection halo -- typically 3.5 arcmin --, the latter is no longer directly  distinguishable. Nonetheless it may still significantly  affect the photometry of the outer regions of the galaxies, in particular falsify their colour. This is particularly evident in Fig.~\ref{fig:halo-color} showing 	the g-r colour map of NGC~3489. Whereas the galaxy has for most of its extent a uniform colour  $g-r = 0.6$, it exhibits an outer annulus with $g-r = 1.1$.
This reddening  is best explained by an  internal instrumentation reflection: it  reflects the red colour  of the galactic nucleus and the fact that the r band is particularly sensitive to such reflections. 
A significant number of the \AD\ galaxies do have a similar reddening which may then be considered as suspicious (Karabal, et al., in prep.).
It badly affects the interpretation of the outer colour gradients in terms of stellar populations.

In fact, such an effect may also be responsible for the reddening of the stellar halos described in a number of published  studies but which is still   controversial  \citep{Bergvall10,Jablonka10,Zackrisson12}. 
In particular the  presence of very red halos on extremely deep images of  some spiral galaxies has generated suggestions about possible unconventional stellar Initial Mass Functions \citep{Bergvall10,Bakos13} but was questioned as they could not be seen in nearby galaxies which are resolved into stars. Our MegaCam images provide an unambiguous  confirmation of the instrumental origin for at least some of the red halos disclosed through their diffuse light. 

Thus a proper study of the outer stellar populations requires to take into account the presence of ghost halos, especially in case  the studied galaxy hosts a bright compact nucleus. 
This requires a physical modeling of the  reflections within the telescope and camera which may be achieved  with ray--tracing experiments (Regnault et al., in prep.).

\subsection{Sky pollution: Galactic cirrus}
\begin{figure*}
\includegraphics[width=\textwidth]{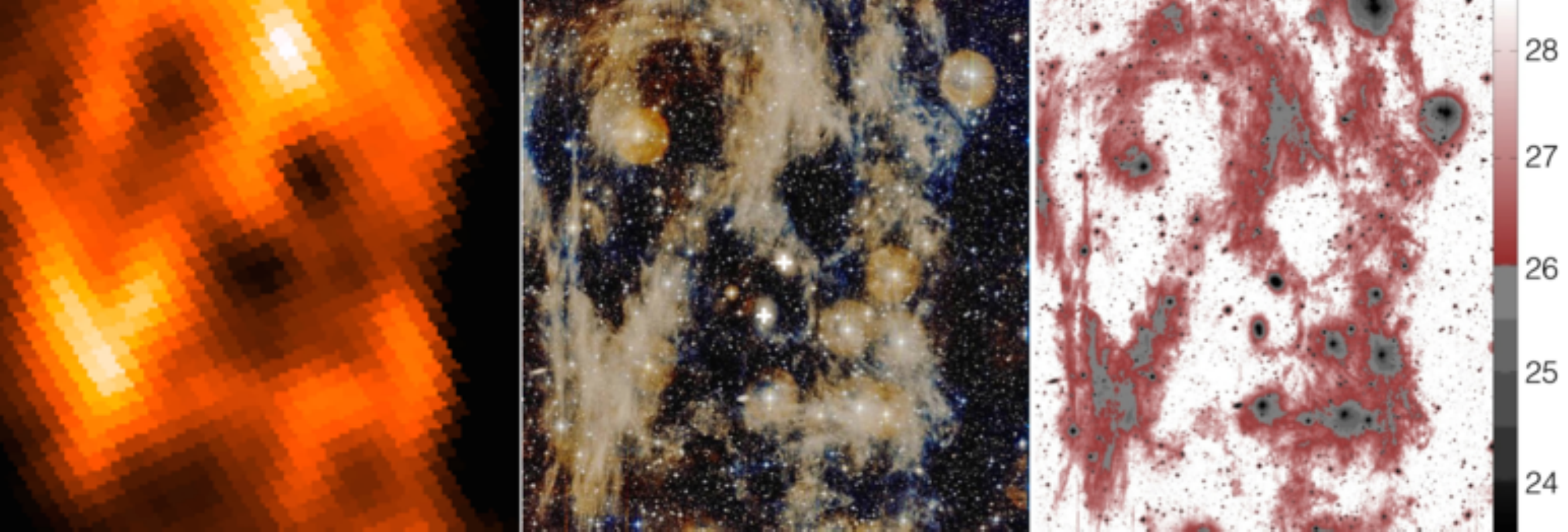}
\caption{Contamination by Galactic cirrus.
{\it Middle:} Composite g+r MegaCam image of the field around NGC~2592 and NGC~2594. The position of each ETG is  indicated by the white cross. The field of view is 51\x57 arcmin. {\it Left:} The same field observed by the HFI camera on board Planck at 857~GHz (350 $\mu$m). Note the very good match between the extended emission in the far IR and the optical, though the MegaCam image has a spatial resolution which is 300 times better than HFI.   {\it Right:} Surface brightness map in the g-band. The scale in \sbr is indicated to the right. Note that most cirrus emission shows up at surface brightness below 26~\sbr (shown in red).   }
\label{fig:cirrus}
\end{figure*}

\begin{figure*}
\includegraphics[width=\textwidth]{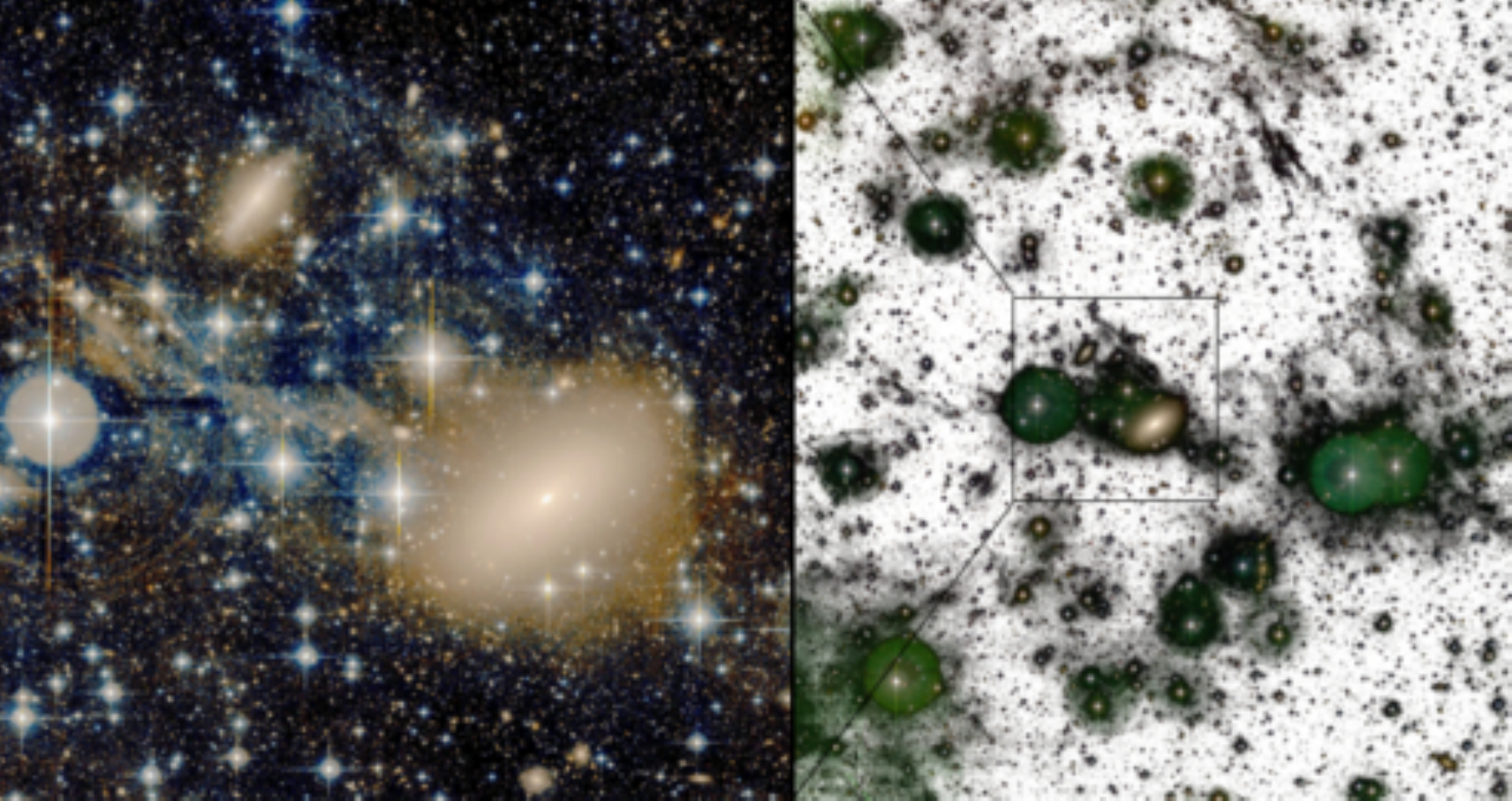}
\caption{A filamentary structure near NGC~7457 resembling  a tidal stellar stream but most likely due to  Galactic cirrus,  as suggested by the zoom out g--band image to the right exhibiting many similar structures all over the field. The  extended halo of the nearby star has been subtracted from the composite g+r image to the left.    }
\label{fig:cirrus-detail}
\end{figure*}

In some fields, the detection of the  low surface brightness stellar structures is  hampered by  the presence of contaminating features covering an even larger area than the reflection halos: Galactic cirrus.
Far infrared/millimetric  emission from cold dust located in the Milky Way is well documented  and appears as one of the principle hurdles in distant galaxy surveys in this wavelength regime   as well as for the study of the Cosmic Microwave Background (CMB). It is less known that Galactic cirrus also affects the optical regime through their scattered light emission, although this light was clearly visible on high-contrast prints of the  Palomar Sky Survey   \cite[e.g.][]{Sandage76}.
At the depth of the MegaCam deep imaging survey, cirrus optical emission becomes prominent.  In the image shown in Fig.~\ref{fig:cirrus}, they occupy about half of the frame. The close match between the diffuse optical structures and the 857 GHz (350~$\mu$m) emission as traced by the HFI camera on board the Planck  telescope leaves no doubt on the origin of these structures.
At the resolution of MegaCam -- 300 times better than HFI --, Galactic cirrus  shows  up as multiple long, narrow filaments sharing the same orientation on the sky over fields of 10--30 arcmin. As seen in Fig.~\ref{fig:cirrus}, cirrus emission becomes prominent only at surface brightness in the the g band fainter than 26~\sbr, and its  colour is globally uniform  though locally colour gradients may be seen. In particular some filaments appear blue on the composite  g+r image (see also Fig.~\ref{fig:cirrus-detail}).

Unfortunately for the purpose of this survey, filamentary structures due to cirrus emission may resemble stellar tidal tails. A particularly striking example is shown on  Fig.~\ref{fig:cirrus-detail}. The long feature emanating East of NGC~7457, which has the same colour as the ETG  halo, resembles that of a stellar stream from a tidally disrupted satellite. Zooming out, one notes however that similar structures are  present all over the field. Thus most likely the feature is a cirrus.  Zooming in, an experienced observer will remark that real stellar streams  are less striped than cirrus filaments. 

Although cirrus is quite easily  identified in most cases, either using complementary multi-wavelength data or examining in detail the MegaCam images, there is no way to subtract it, like it is done for instance to generate CMB maps, without at the same time erasing the real LSB stellar structures.
About 30~\% of the MegaCam images in our survey exhibit emission  from cirrus. For about 15~\% of them, the contamination is so strong that the detection of low surface brightness stellar features becomes quasi impossible. The degree of contamination depends on the galaxy latitude and may be predicted from the Planck emission at 857~GHz. For that purpose we have extracted from the Planck archives the area covered by the MegaCam images and measured within them the average flux (mostly dominated by foreground cirrus emission).
In fields with  a mean  857~GHz flux exceeding  1.5 MJy/sr, 100~\% of MegaCam images are contaminated. Between 1 and 1.2  MJy/sr, the contamination level is 50~\%. It is only below fluxes of 0.4 MJy/sr, that the contamination becomes less than 10~\%. 

Whereas the optical emission of Galactic cirrus is a stumbling block for extragalactic research, it may be of great interest for the study of the interstellar medium. Indeed,   the MegaCam images provide us with maps of Galactic cirrus at the unprecedented spatial resolution of about 1 arcsec whereas direct images in the FIR, for instance with Planck, have a resolution 300 times worse. A study dedicated to the cirrus emission will be published in a separate paper.

\subsection{Surface brightness limit and comparison with other surveys}
\label{sec:limit}

A key parameter for any deep imaging survey is the limiting surface brightness. Unfortunately its value is much more difficult to estimate than the limiting magnitude of traditional imaging surveys obtained with aperture photometry of point-like sources. This is largely due to the fact that the limiting factors are the systematics in the  background. 
The large range of limiting surface brightness quoted in the literature for LSB optimized surveys does not  only reflect  the range of exposure times, telescope size used and real achieved depth, but also  the diversity of methods used to do the photometry. 

To set the scene, the SDSS --  the current reference for imaging survey -- has a limiting surface brightness in the g--band of 26.4~\sbr \citep{York00,Kniazev04}, with a  gain stacking the g, r and i bands \citep{Miskolczi11}, or for fields with repeated observations such as in the Stripe~82 calibration area \citep{Kim13}.

\noindent $\bullet$  \cite{Atkinson13} used the  CFHTLS-Wide imaging survey  (which was not LSB optimized)   to probe  tidal features around galaxies within the redshift range 0.04 -- 0.2. They estimated a limiting surface brightness of   27.7~$\pm$ 0.5 ~\sbr in the g--band based on histograms of rms  variations of the sky noise estimated in multiple regions.\\
\noindent $\bullet$    \cite{Tal09} used pointed observations with the SMARTS 1m telescope to determine the frequency of tidal features in a sample of 55 elliptical galaxies. They  heavily smoothed  dark-sky and target images to  the scale of typical tidal features  and determined the 1$\sigma$ detection threshold of the latter. 
They obtained  a limiting surface brightness of  29~\sbr in the V(Vega)--band -- , and 27.7 (about 27.9 in the MegaCam g--band, and AB scale), using flatness-limited frames.\\
\noindent $\bullet$  The pilot survey of \cite{Martinez-Delgado10} of tidal streams around nearby spirals done with a luminance (L) broad  filter and robotic amateur telescopes could not be directly   photometrically calibrated, but using as a reference SDSS images, the authors estimated from the background fluctuation an equivalent V/g--band limiting magnitude of  28.5~$\pm$ 0.5~\sbr.  \\
\noindent $\bullet$  
The LSB--optimized NGVS at the CFHT \citep{Ferrarese12}  reaches  29~\sbr in the g--band, a value checked comparing the surface brightness profile of the galaxy M49 with previously determined ones  in the literature. \\
\noindent $\bullet$ 
 \cite{Bridge10} used the deep component of the CFHTLS to estimate the merger rate of galaxies up to a redshift of 0.2. They claim a very low limiting surface brightness in the i--band of 29~\sbr  (thus about 30~\sbr in g), but do not state in their paper the method used to estimate it.  \\
 \noindent $\bullet$ 
\cite{Sheen12} probed post-merger signatures in nearby cluster galaxies on images obtained with the Blanco 4m telescope supposedly reaching 30~\sbr in r (about 30.7 in g).\\ 
\noindent $\bullet$ 
Finally\footnote{A more comprehensive census of previous imaging surveys may be found in Table~1 of \cite{Atkinson13}.}, 
\cite{vanDokkum14} testing his LSB-dedicated camera -- the Dragonfly Telephoto Array -- on the spiral  M101 holds the current record of  claimed  limiting surface brightness with an amazing value of 32~\sbr in g,   estimated looking at the shape and extent of the galaxy radial profile.  If confirmed,  such a depth would reach that obtained with star counts in the Local Group, offering a tremendous opening for galactic archeology in the nearby Universe. 

How does our survey compare with previous ones? As explained in Sect.~\ref{sec:obs}, the Elixir-LSB pipeline achieves  a sky flattening  of 0.2\%,  nearly 7 magnitudes fainter than the sky background.
We estimated a nominal detection limit at 28.5~\sbr in g, determining  the pixel value  difference between  the bump and holes on  clean sky regions.
Whether this nominal value is realistic can be checked looking at the 2D surface brightness maps and 1D profile of real galaxies. 

\begin{figure*}
\includegraphics[width=\textwidth]{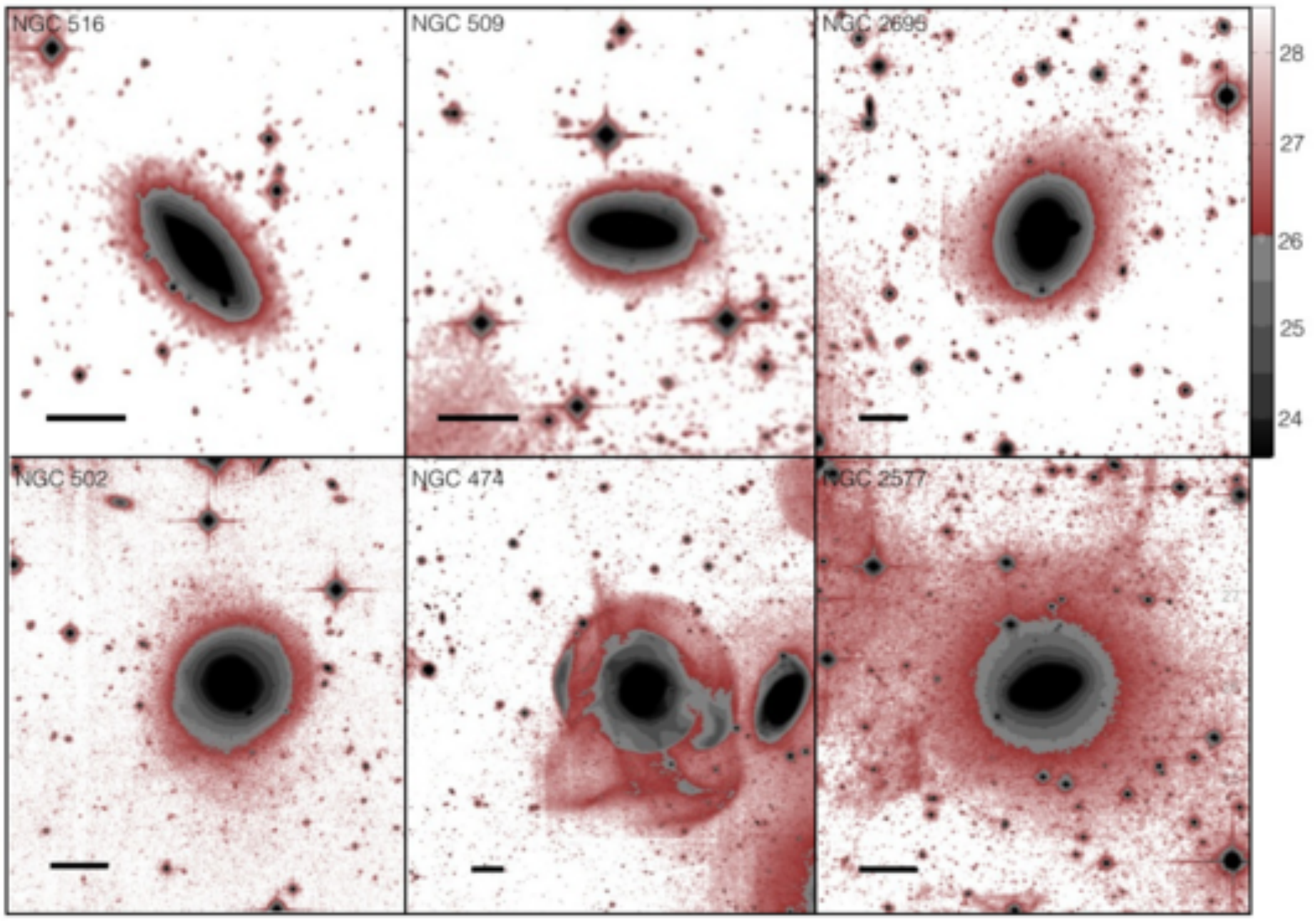}
\caption{Surface brightness maps in the g-band of several ETGs. The brightness scale used, shown at the top--right -- ladder-type in grey scale below 26~\sbr, linear and red above --, enhances the low surface brightness structures disclosed by the deep MegaCam images. The galaxies, resp.  NGC~516,  NGC~509, NGC~2695, NGC~502, NGC~474 and NGC~2577  are ordered by relative increased contribution of this LSB component to the total light. 26~\sbr\ roughly corresponds to the surface brightness limit of regular imaging surveys such as the SDSS. Each bar corresponds to 10~kpc at the distance of the galaxy.}
\label{fig:sb}
\end{figure*}

Examples of surface brightness maps are shown in Fig.~\ref{fig:sb} with the colours of the  intensity scale chosen to highlight structures fainter than  26~\sbr, i.e. structures that are not typically seen in regular imaging surveys such as the SDSS.
In these images, all features until at least 28.5~\sbr are ``real": they do not have the shape of instrumental signatures (see above). This gives an upper limit to our limiting surface brightness sensitivity.
Surface brightness profiles will be discussed in another paper.

\section{Data products}
\label{sec:prod}

\subsection{Sample presented in this paper and selection biases}
\label{sec:sample}
The \AD\ sample is  volume limited: it includes galaxies located at a distance below 42~Mpc, classified as early-type based on a visual classification criterion -- the absence of discs and prominent dust lanes on shallow images --, with absolute K--band magnitude brighter than  -21.5 (i.e. a stellar mass above $6 \x 10^{9}~\Mo$) -- thus dwarf ETG galaxies are excluded --, and  further restrictions on declination -- the galaxies should be  visible from northern facilities -- and Galactic latitude \citep[see details in][]{Cappellari11}.
All galaxies in the \AD\ sample benefit from a wealth of ancillary multi-wavelength data, including the internal stellar kinematics derived from SAURON observations, based on which they have been classified as slow or fast rotators \citep{Emsellem11}. 

We present here an image catalog for a sub--sample  of  92 galaxies which have been observed with MegaCam through at least the g and r filters. Among those 59~\%, have also  i--band images and 9~\%, u--band images.

\begin{figure}
\includegraphics[angle=-90,width=\columnwidth]{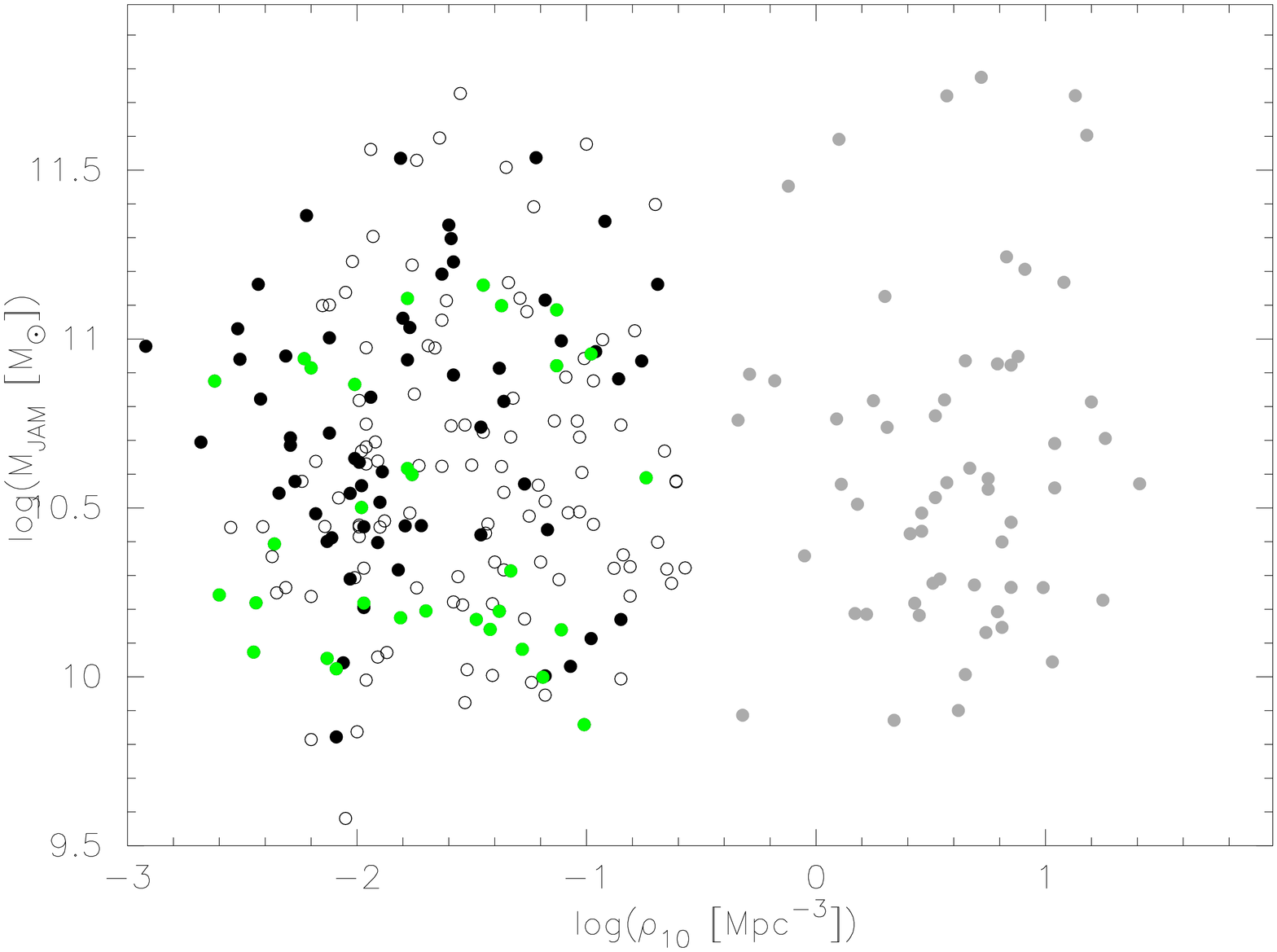}
\caption{Selected ETGs in this catalog plotted on a diagram showing the dynamical mass, M$_{\rm JAM}$, derived from  modeling of the stellar kinematics and the isophotal structure of each galaxy,  versus the volume density diagram, $\rho_{\rm 10}$, in log (Mpc$^{-3}$).  
Open circles correspond to the full \AD\ sample, and filled discs are galaxies with available deep imaging. Note that all galaxies located in regions with density above log($\rho_{\rm 10})=-0.4$ belong to the Virgo Cluster and were observed as part of the NGVS project (filled grey discs). They will be presented in a separate paper. Relaxed galaxies showing no sign of tidal perturbations even with the deep imaging are shown in green.}
\label{fig:loc}
\end{figure}

\begin{figure}
\includegraphics[angle=-90,width=\columnwidth]{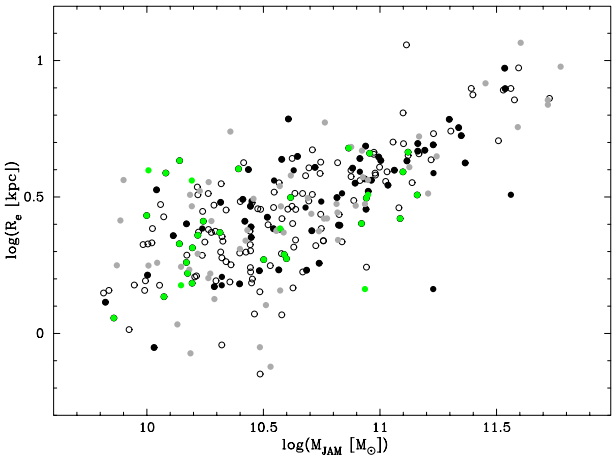}
\caption{Selected ETGs in this catalog plotted on a diagram showing the dynamical mass, M$_{\rm JAM}$, derived from  modeling of the stellar kinematics and the isophotal structure of each galaxy  versus  the effective radius, \Reff\  \citep[see][]{Cappellari13}.    Open circles correspond to the full \AD\ sample, and filled discs are galaxies with available deep imaging (grey ones belong to Virgo and were not analyzed here). Relaxed galaxies showing no sign of tidal perturbations even with the deep imaging are shown in green. }
\label{fig:MRe}
\end{figure}

\begin{figure}
\includegraphics[angle=-90,width=\columnwidth]{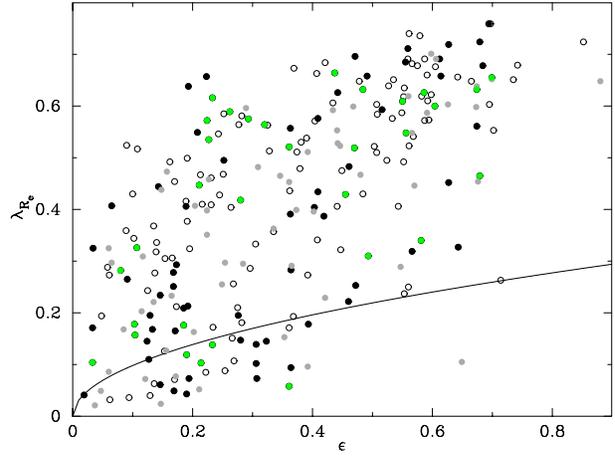}
\caption{Selected ETGs in this catalog plotted on a diagram showing the specific angular momentum  at the effective radius, $\lambda_{\rm R_e}$, versus the ellipticity, $\epsilon$.  The solid line divides the fast  (top) and slow (bottom) rotators according to the \AD classification scheme  \citep{Emsellem11}.  Open circles correspond to the full \AD\ sample, and filled discs are galaxies with available deep imaging (grey ones belong to Virgo and were not analyzed here). Relaxed galaxies showing no sign of tidal perturbations even with the deep imaging are shown in green. }
\label{fig:rot}
\end{figure}

The  ETGs presented in this paper  are located in environments with low to medium galaxy density, excluding the Virgo Cluster. 
This is illustrated in Fig.~\ref{fig:loc} showing  for the entire \AD\ sample and the sub-sample discussed here the stellar mass versus  $\rho_{\rm 10}$, i.e.  the volume density in Mpc$^{3}$ of  galaxies inside a sphere of radius r$_{\rm 10}$ centred on a galaxy, which includes N$_{\rm gal}=10$ nearest neighbours \citep{Cappellari11b}.
This sub-sample of 92 galaxies -- already by far the largest sample of early-type galaxies with available deep images -- spans all the \AD\ range of masses, but high mass and extended galaxies are over-represented (see Fig.~\ref{fig:MRe}). As a consequence of the selection criteria used for the  MegaCam runs, a large fraction  of the  rather rare slow rotators have  deep images, while the numerous fast rotators are much less sampled (see Fig.~\ref{fig:rot}). Finally, the sample is also biased towards gas--rich objects, i.e. ETGs for which \HI\ and CO line emission has been detected.

The  CFHT Large program MATLAS  aims at eluding such selection effects. When completed,  this survey done with the same observing strategy as the one described  here, will  provide deep images at similar depth for all 260 \AD\ galaxies. 
Note that the  ETGs  located in the Virgo Cluster -- about one fourth of the \AD\ sample --    were already observed with MegaCam  as part of the  CFHT Next Generation Virgo Cluster Survey \citep[NGVS,][]{Ferrarese12}.  Their images will be published in the framework of the NGVS collaboration (Duc et al., in prep.). 

Consequently  this paper  focuses  on general issues faced by deep images and addresses  effects which are not influenced by selection biases while the in-depth analysis  statistical analysis is postponed to future papers.

\subsection{High level image production}

\subsubsection{True colour  images}
True colour images obtained combining multi-wavelength bands have not only an aesthetic value\footnote{Images from the MegaCam survey have appeared in calendars, in an art gallery  and the Astronomical Picture of the Day (see http://apod.nasa.gov/apod/ap140105.html )}, they are particularly efficient at synthesizing information. 
Depending on the filters available,  the composite images were either produced using    the g (blue channel), r (green channel) and i (red channel) or only with the g and r bands. In the absence of a third band, a fair true colour rendering   may be obtained using as the middle green  channel the combination of the blue and red image (here g and r). An arcsinh intensity scale was applied to each channel in order to decrease the dynamical scale and make visible both the central and outer regions. The same weighting for the red, green and blue channels was applied to each galaxy, enabling a qualitative comparison of  the colours between objects observed with the same filter set. Final image combination was carried out with {\sc STIFF} \citep{Bertin11}.

\subsubsection{Surface brightness maps}
Surface brightness maps with an intensity scale in ~\sbr\  are computed on sky--subtracted images. As the Elixir-LSB process already produces flat images, a simple constant was subtracted.
To compute the average sky value, a histogram of the pixel values over the full MegaCam stacked image  was determined and the mode of the distribution was used. Note that with this method, the sky level might have been overestimated in frames contaminated by extended cirrus emission or full of bright stars. 
In order to increase the contrast of LSB structures, an adaptive smoothing algorithm was applied to the data, using the software {\sc ADAPTMOOTH} of \cite{Zibetti10}. 
Pixels are grouped together and averaged to keep the same S/N over the whole image. Contrary to regular smoothing techniques, the  spatial scale of bright objects, i.e. the central regions of galaxies, stars, etc., is preserved with this scheme.

\subsubsection{Residual images}
\label{sec:model}
Historically, the presence of fine structures in  galaxies, in particular ripples and shells, was discovered within their diffuse stellar halos, and was disclosed with various techniques of contrast enhancement such as  unsharp masking \citep{Malin77,Malin83,Schweizer88}.   
Our sensitivity enables the direct detection of LSB structures well outside the outer galactic halos. Removing the latter allows us however to connect the outer and inner fine structures and make their complete census. 
The galaxy modeling required for this was done with two techniques: a multi-component parametrization of the host galaxy with {\sc GALFIT} \citep{Peng02}, and  ellipse fitting using the eponymous  package within  the {\sc IRAF} software \citep{Jedrzejewski87}.   We present in this catalog  residual images obtained subtracting galaxies modeled by the ellipse fitting algorithm.
Both techniques generally give similar results, except in the very central regions, for which residual images are less noisy with GALFIT (as it may model non  axisymmetric  components), and in the very outer regions, best subtracted with the ellipse model. As the focus of our project is these external regions, we made the choice of preferentially using the  ellipse models in our analysis.

As a first step, point-like objects in the field were identified with {\sc SExtractor}  \citep{Bertin96}. Faint stars were removed replacing them with the local background determined from surrounding pixels. The  extended reflection  halos due to the bright nearby stars have been subtracted with the technique described in Sect.~\ref{sec:halos}. Remaining extended objects not associated to the central ETG, like bright stars, background or companion galaxies, were masked. The ellipse fitting was then performed, leaving whenever possible the central position, axis ratio and angle parameters  free. In a few cases, those parameters had to be kept constant to avoid  divergence. The resulting ellipse model was then subtracted from the image. The process was  iterated  to remove additional stars that were not identified in the original image. 

\begin{figure*}
\includegraphics[width=\textwidth]{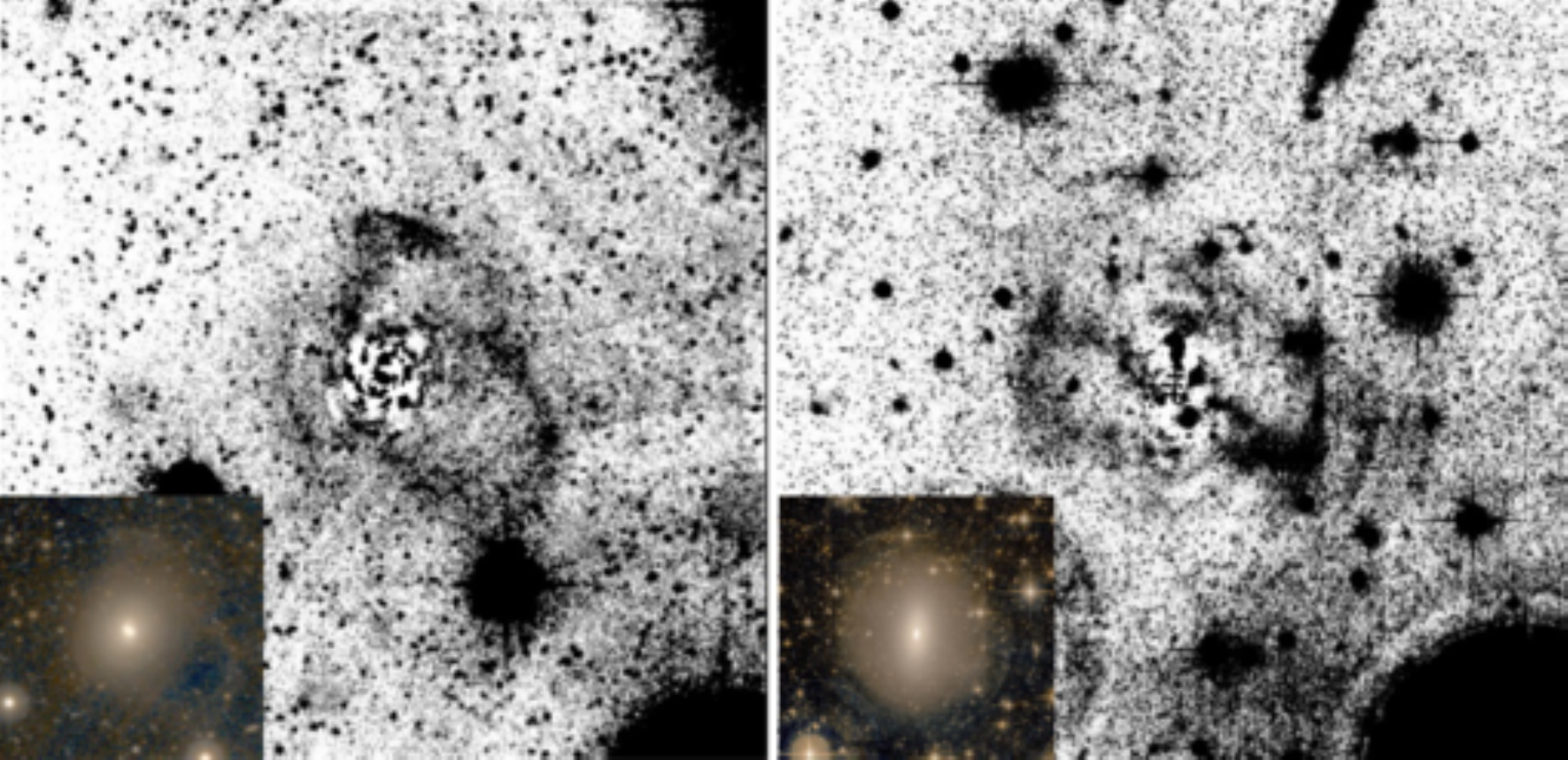}
\caption{Examples of fine structures being revealed after subtracting a galaxy model made with ellipse fitting.
Residual g--band image of UGC~05408 (left) and NGC~3245 (right). The insets show  true colour g+r images of the central regions  before the galaxy subtraction. }
\label{fig:ellipse}
\end{figure*}

Fig.~\ref{fig:ellipse} presents two examples of  images with the  host galaxy subtracted. Such residual images  are especially useful to reveal the shape of  fine structures that were barely  visible  on the original images. 

\subsection{Image catalog}
\label{sec:cat}

\begin{figure*}
\includegraphics[width=\textwidth]{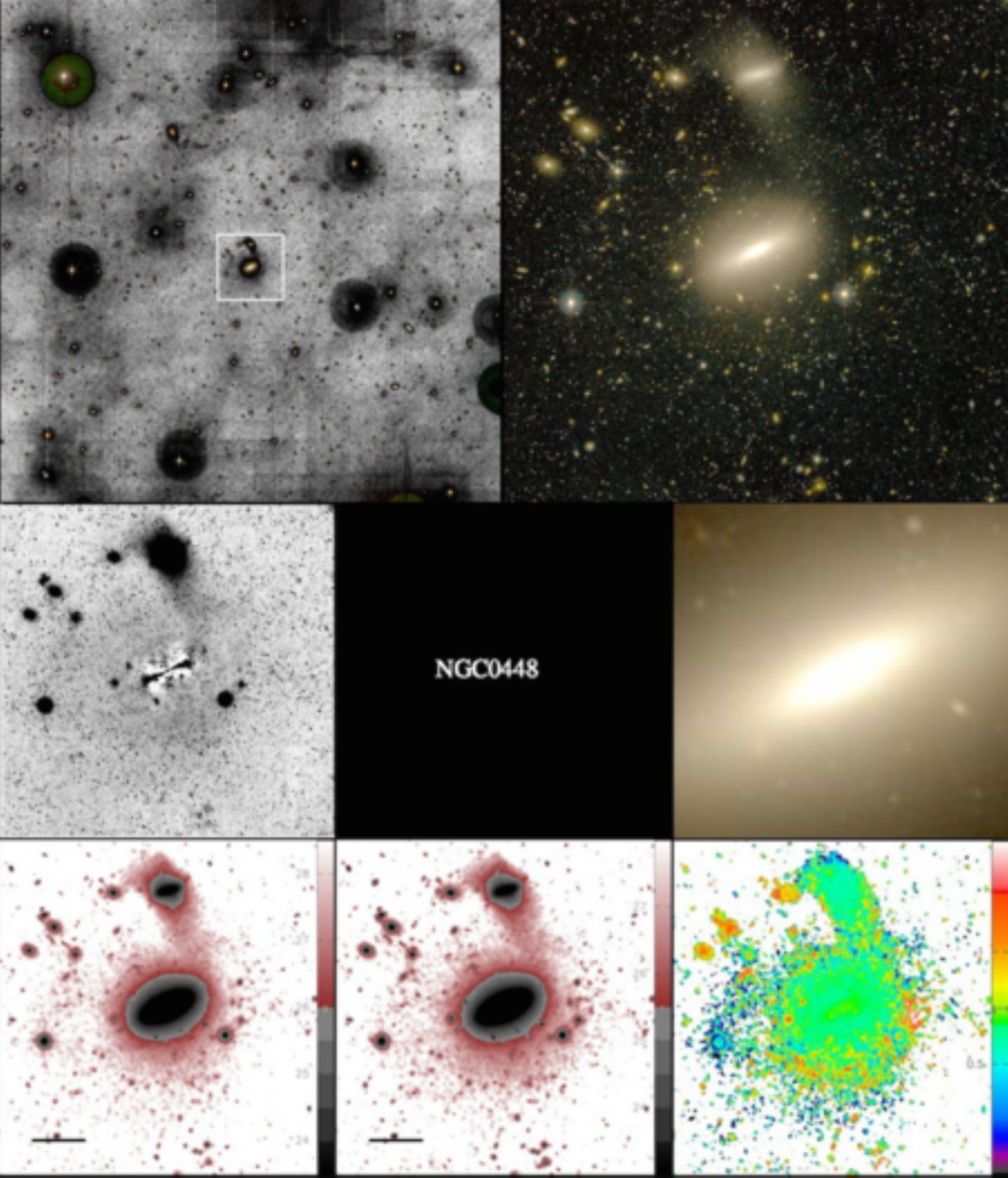}
\caption{Set of images for the first entry in the on--line catalog, NGC~448. See  Sect.~\ref{sec:cat} in main text for a detailed description of each panel.  }
\label{fig:cat}
\end{figure*}

We present in this catalog images of  92 ETGs. Derived products, such as the surface brightness and colour profiles, will be published elsewhere.
The full atlas, only available in electronic form (see Fig.~\ref{fig:cat} for the first galaxy in the catalog)\footnote{The true colour images, surface brightness and colour maps, image residuals may also be explored on line  with the {\sc Google} navigation tool API  (including zooming in/out and quick channel change capabilities). Access to the database is  available through the \AD\ site http://purl.org/atlas3d. 
Raw fits data are available from the Canadian Astronomy Data Centre (CADC) which hosts the CFHT archives.},  consists of a series of images:\\ 

\begin{itemize}
\item {\it Top left:}  g--band image of the original MegaCam stacked frame hosting the ETG. It is shown in grey linear scale with a small cut range so as to increase the contrast of low surface brightness features. The displayed field of view is 57.0 \x 57.0 arcmin.
To identify the underlying objects -- stars and galaxies --, a composite g+r or g+r+i image  has been superimposed. 
The target galaxy is indicated by a white square. Its size corresponds to 40 times the effective radius of the galaxy, as determined in \cite{Cappellari11}. This delineates the area shown in all the other sub-panels in the figure, except that zooming into the central regions. 

\item  {\it Top right:} a  ``true colour" (g+r+i, or g+r) image centered on the galaxy.  Contrary to some true-colour images shown throughout the paper,  no local correction was done on these images, and the signature of instrumental artefacts was kept. 

\item  {\it Middle  right:} same true colour image zoomed on the central regions by a factor of 6. 

\item  {\it Bottom left:} g--band surface brightness map.
Between 23.5 and 26~\sbr\, the intensity scale used -- a ladder scale, with step of 0.5~mag and grey colours -- discloses the global elliptical shape of the ETG. Above 26~\sbr\, the intensity scale is linear and a red colour is used to highlight the new structures revealed by the deep MegaCam survey. The horizontal bar corresponds to 10~kpc at the distance of the galaxy.

\item  {\it Bottom middle:} idem but for the r--band. The intensity scale has been shifted by 0.5~mag to take into account  the colour of the galaxy. 

\item  {\it Bottom right:} corresponding g-r colour map. The colour bar, here displayed with a linear scale, is indicated to the right. Pixels below the detection limit are shown in white.

\item  {\it Middle left:} residual image obtained subtracting from the g--band a galaxy model made with the ellipse fitting procedure described in Sect.~\ref{sec:model}

\end{itemize}
In all maps, North is up and East left.\\

\section{(Re)-classifying galaxies with deep imaging}
\label{sec:class}

Galaxy classification is now often  done based on:\\
-- their global colour: galaxies may belong  to the red sequence, green valley or blue cloud.\\
-- ability or not to form stars: galaxies are passive, quiescent, active, star-bursting.\\
-- internal kinematics: galaxies are fast or slow rotators.\\
The kinematics criteria is at the base of the classification scheme proposed by the Sauron survey \citep{Emsellem07} and refined by the \AD\ project \citep{Cappellari11b}, which perhaps tells more about their origin and evolution than other properties.
  However, apparent morphologies -- the importance of the bulge vs disc component, the presence or absence of spiral arms and bars, the degree of tidal perturbation  -- remain an unavoidable criterion in the taxonomy of galaxies. 

The Hubble sequence likely gives a misleading view of the diversity of galaxies but continues to be widely used, even at high redshift.
In that respect although astronomy has for long entered in a multi-wavelength  area, optical imaging remains key. In times when the credibility of any discovery relies  on its assessment on large samples -- especially in the nearby Universe --, galaxy classification relies on wide but relatively shallow surveys, such as the SDSS. The various galaxy zoo projects \citep{Lintott08} involving  hundreds of thousands of citizen scientists inspecting millions of SDSS images   has been instrumental in the classification of galaxies. 
However, how much  were these tremendous efforts impacted by the use of images reaching a limited surface brightness? Our  deep imaging survey of already a substantial number of galaxies may address any putative bias. Initial results  are presented here. 

\subsection{Global morphologies}

We first address the global appearance of the galaxies imaged by MegaCam. The 260 Early-Type Galaxies in the \AD\ sample were initially selected after a visual inspection of a volume limited parent sample of 871 galaxies of all types. Following the classic criteria which define the revised Hubble classification scheme outlined by \cite{Sandage61}, those lacking spiral structure or prominent dust lanes were considered as ETGs. This selection was  done based on SDSS images, whenever available, and images from the Digital Sky Survey (DSS). The classification  was further checked with images  acquired at the Isaac Newton Telescope having the same depth as the SDSS ones. Would the result have been the same if instead the much deeper MegaCam images had been used?
We re-examine here the principle components that supposedly disentangle spirals from lenticulars and ellipticals, but also relaxed, evolved systems  from perturbed ones that do not fit on the Hubble sequence.

\subsubsection{Presence of spiral and ring structures}

\begin{figure*}
\includegraphics[width=\textwidth]{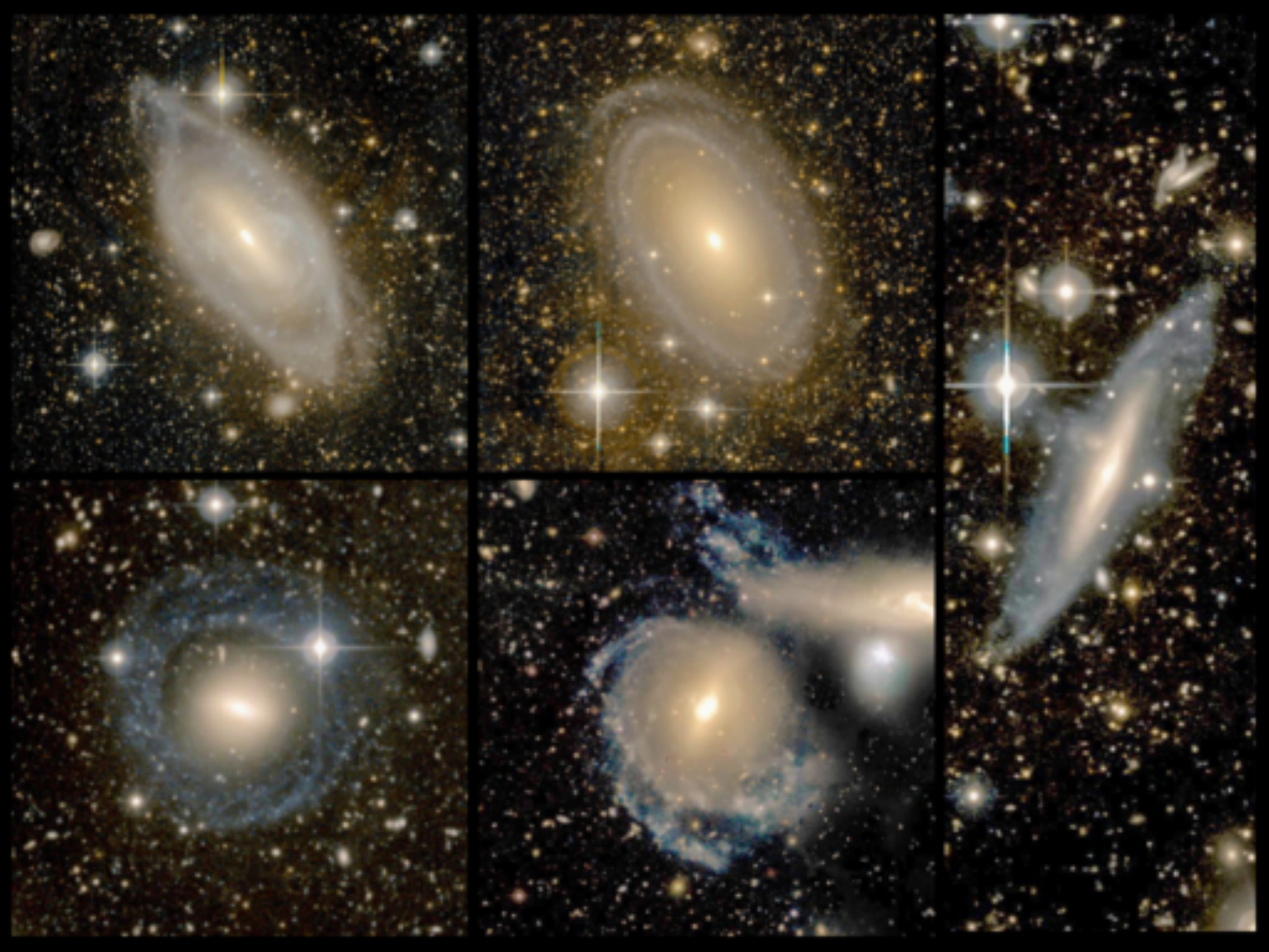}
\caption{Examples of ETGs  for which deep optical images reveal the presence of low surface brightness, blue, star-forming discs, rings or spiral structures around them. Composite true colour g+r+i i or g+r images of (clockwise) NGC~2685, NGC~5582,  PGC~016060, NGC~7465 and UGC~09519. Field of views and physical scales which vary from one galaxy to the other  may be found in the image catalog. Manual local corrections erasing instrumental signatures and bright halos as well as colour enhancements were performed  for a better display. Unprocessed images can be inspected in the catalog. }
\label{fig:disk}
\end{figure*}

\begin{figure}
\includegraphics[width=\columnwidth]{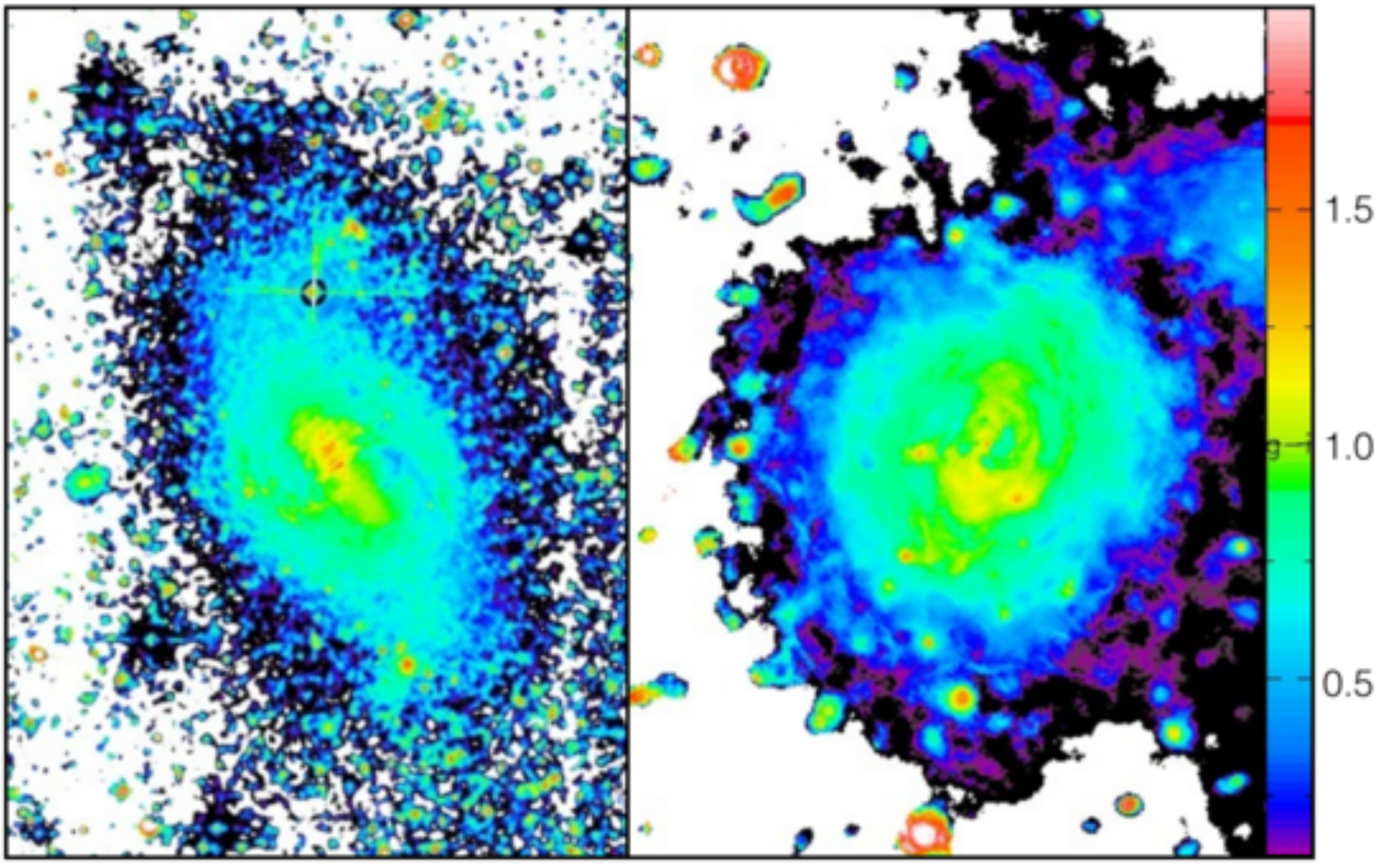}
\caption{g-i colour maps of two ETGs with blue discs around them: NGC~2685 (left) and  NGC~7465 (right). The colour scale in \sbr\ is indicated to the right.}
\label{fig:disk-col}
\end{figure}

Fig.~\ref{fig:disk} presents composite true colour images of 5 ``ETGs'', for which MegaCam revealed the presence of an extended blue low surface brightness component surrounding the main (reddish) body. Spiral arms are visible in these structures which are presumably star--forming. Among those only NGC~2685, a polar--ring galaxy nicknamed Spindle or Helix, was already known to host an outer stellar ring \citep[e.g.][]{Sarzi06,Jozsa09}. Note that  NGC~5582,  classified as  ``E''  in the RC3 catalog,  would have been classified as a spiral galaxy based on our deep photometry alone, if one considers the disc component extending well outside the galaxy optical radius.
The presence of star--forming regions in the outskirts of some ETGs was already known from UV observations \citep[e.g.][]{Jeong07,Donovan09,Marino11}, with possibly \HI\ fueling them  \citep[e.g.][]{vanDriel91,Serra12}. However, their  optical counterpart had so far been elusive. 
Besides, the previously UV--luminous rings documented  in the literature \citep{Salim12} are most often located within the main body of the galaxy, contrary to those revealed by MegaCam located at typically 5 $\Reff$. 
In the examples shown on  Fig.~\ref{fig:disk} (see in particular UGC~09519), the ``red and dead" component of the ETG is spatially separated from the more active, blue one. This is best seen on the colour maps of these galaxies shown on  Fig.~\ref{fig:disk-col}, showing an abrupt blueing of the colour profile at radial distances above 10~kpc.

The possible external origin of these apparently decoupled structures will be investigated coupling  \HI\  \citep{Serra12}, CO \citep{Davis11} and optical observations.

\subsubsection{Inner disturbances: merger remnants}
\label{sec:mergers}
A  scenario for the formation of massive ellipticals    through the merger of  spiral galaxies was proposed decades ago after the first numerical simulations of collisions had been performed \citep{Toomre77a,Schweizer82,Barnes92}, and has since been  a passionate subject of discussion. In any case,  if this hypothesis is correct, at least a fraction of ETGs should be post-mergers, and exhibit their emblematic traces: tidally disturbed morphology and prominent  dust lanes, also revealing the presence of accreted gas. 

\begin{figure*}
\includegraphics[width=\textwidth]{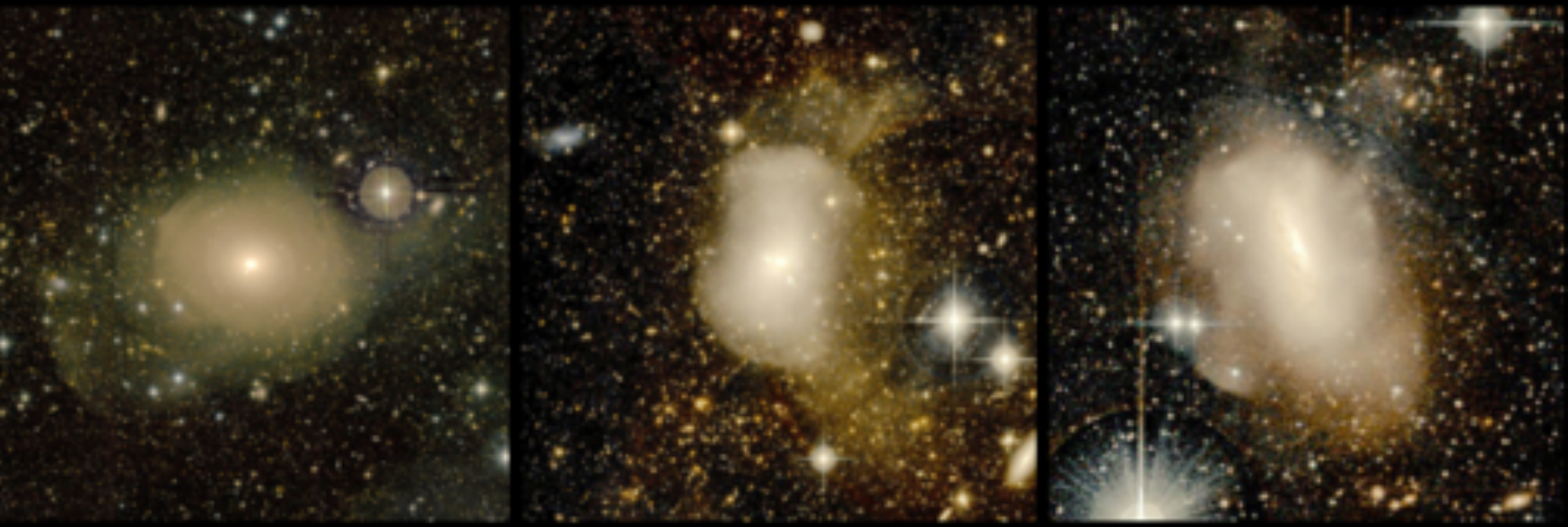}
\caption{Examples of  ETGs with  post- major merger signatures:  inner dust lanes, strongly perturbed morphology and prominent tidal tails. From left to right, composite true colour images of NGC~5557, NGC~1222 and NGC~2764.} 
\label{fig:mergers}
\end{figure*}

Fig.~\ref{fig:mergers} displays 3 examples of ETGs showing on the MegaCam images unambiguous signs of a past major merger event. The dusty ETGs NGC~1222 and NGC~2764 were already classified as peculiar in the NASA/IPAC Extragalactic Database. This was not the case for NGC~5557, classified as an E1 in the RC3 catalog: its post-merger nature revealed by the deep imaging was discussed in \cite{Duc11}.

We have presented here cases of galaxies presenting a {\it global} disturbed morphology, indicative of a (relatively) recent merger. Many apparently relaxed  ETGS in the sample exhibit around them   fine structures, including tidal tails, which  likely trace on-going interactions, but for some of them could be the vestiges of old collisions. They are addressed in Sect.~\ref{sec:FS}.

\subsubsection{Relaxed systems}
\begin{figure*}
\includegraphics[width=\textwidth]{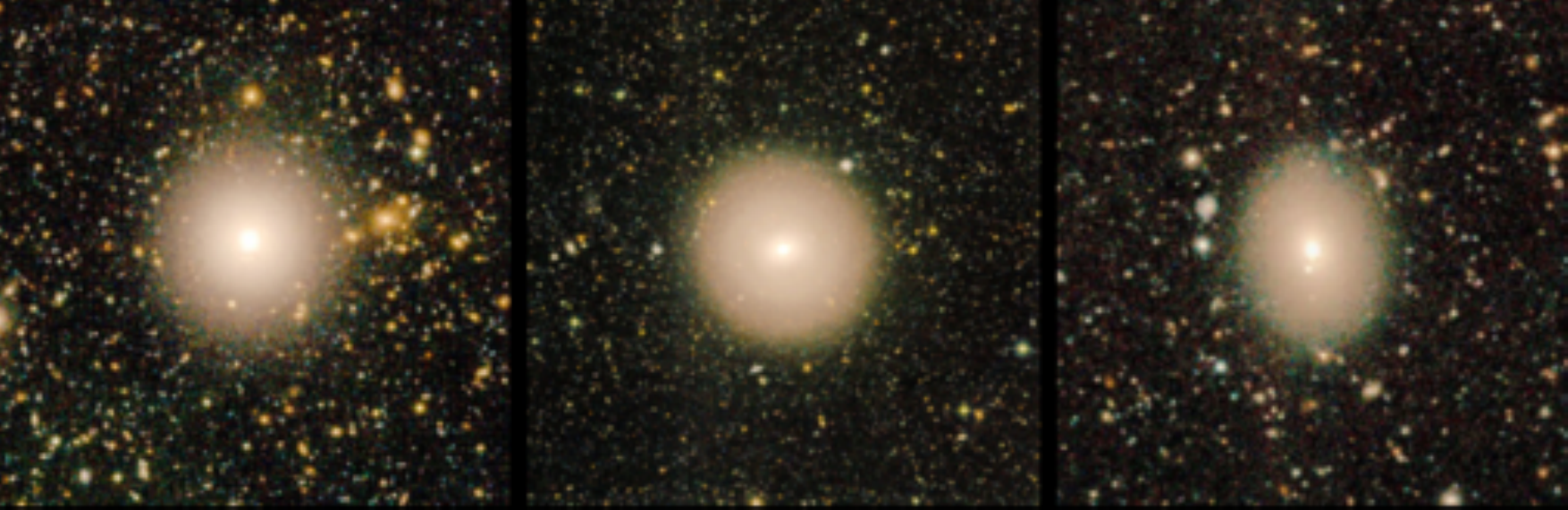}
\caption{Examples of ETGs which appear totally relaxed/regular even on the deep MegaCam images, and do not shown any fine structure in their vicinity. From left to right, composite true colour images of NGC~3457, NGC~3599 and PGC050395. The high Image Quality of the MegaCam survey is illustrated by these images showing the globular cluster population around the galaxies and  background distant clusters of galaxies.}
\label{fig:relaxed}
\end{figure*}

The  images presented so far in this paper may  give a misleading impression of this deep imaging exercise. Not all galaxies in our sample exhibit external star-forming discs, tidal tails and shells. Even at the depth of the MegaCam survey, a large fraction  of them remain the featureless ``red and dead" galaxies that they were thought to be. The  images of three of them are shown in Fig.~\ref{fig:relaxed}.

Cosmological simulations made in the framework of the hierarchical model of the Universe predict that each galaxy is the result of multiple minor and major mergers and should thus be surrounded by vestiges of such collisions \cite[e.g][]{Bullock05,Helmi11}. This should be even more the case for ellipticals as they are believed to be the end--product of the mass assembly process at the galactic scale. Determining the fraction of ETGs that do not exhibit such collisional debris, and understanding why --  a property specific to a sub--class of old ETGs,  a lack of sensitivity or issues in the models --   are among the major goals of our survey.

\subsection{Fine structures}
\label{sec:FS}

The  prime original motivation of  most deep imaging surveys  is the detection of  the so--called ``fine structures". The literature is rather elusive in the definition of this class of astronomical objects which gathers tails, streams, shells, ripples, etc. Authors tend to use different names for the same physical objects or reversely use the same  name for  objects that have a different intrinsic nature or origin. This is especially problematic for their census and classification, and later on use as archeological probes 
\citep[see the  reviews on collisional debris by][]{Duc13,Duc13b}. 
The following section is an attempt to give a  phenomenological description of each type of fine structure and provide a unique name for each different physical class. The ambition is not only to determine the fraction of galaxies having merger vestiges in their vicinity, as usually done, but to make the census of fine structures  and distribute them within different sub--classes, as each of them is a probe of different past events: major, minor, wet or dry merger. 
Obviously when confronted with noisy images, making an unambiguous  distinction between what we later call a tail or a stream may be considered hopeless.  However the intrinsic errors in such  exercice may at least be compensated by the large number of objects and of volunteers making the classification.  

\subsubsection{Tidal tails}

\begin{figure*}
\includegraphics[width=\textwidth]{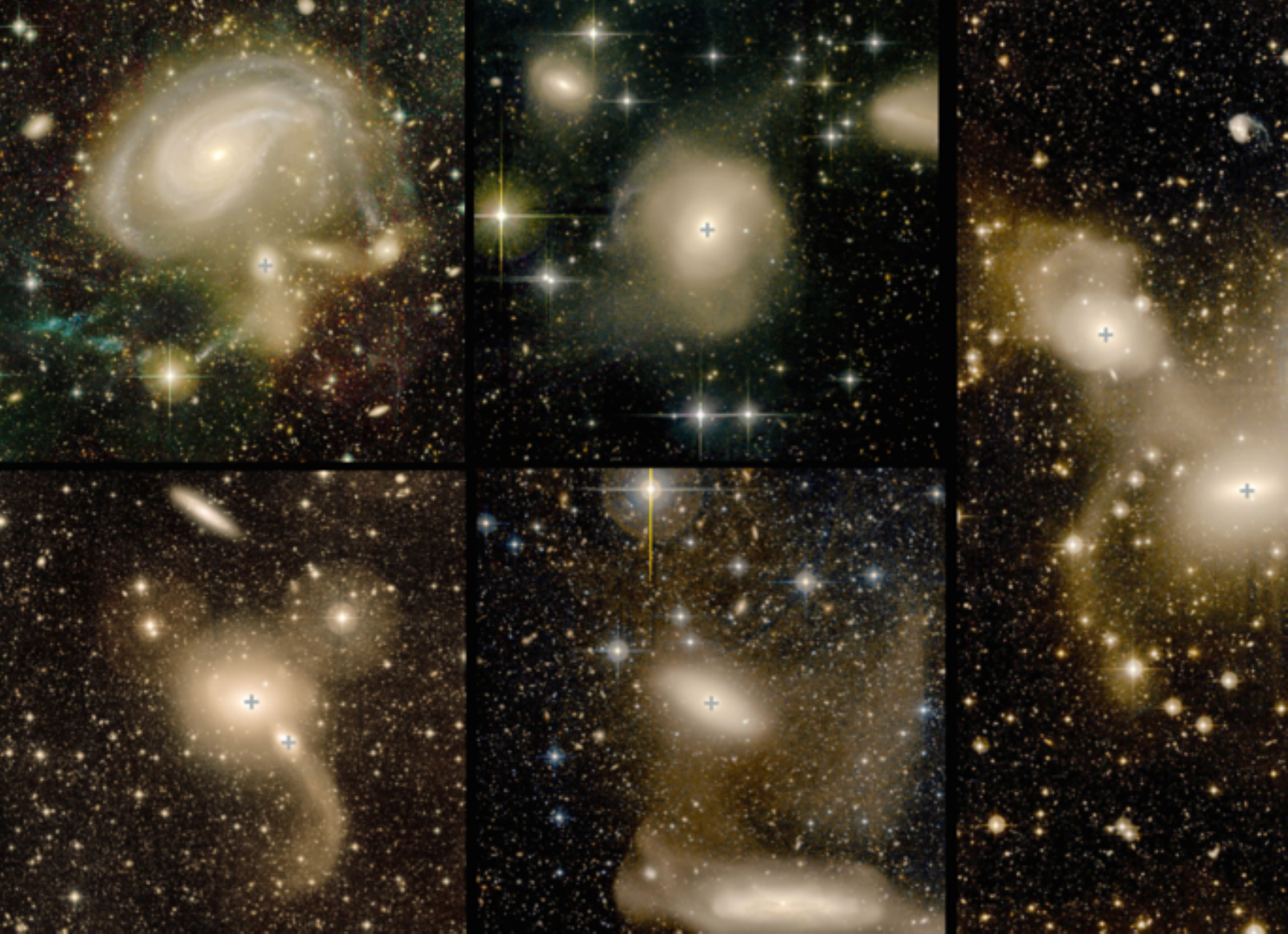}
\caption{Examples of ETGs currently involved in a tidal interaction with a nearby massive  companion  and exhibiting prominent tidal tails. Clockwise, composite true colour images of NGC~770,  NGC~680, NGC~2698/99, NGC~5507 and NGC~5574/76. The location of the \AD\ ETGs is indicated by a cross.}
\label{fig:tails}
\end{figure*}

Tidal tails are the generic names of the elongated structures shaped by any tidal interaction. We restrict here  the definition  to the structures solely made during major mergers. They consist of material  expelled from the primary galaxy, following an interaction with a companion massive enough to have significantly perturbed it. As a consequence, the outer regions of the primary -- here the ETG -- and the tails emanating from them share the same properties. This means similar age and metallicity,  translating to similar colours, for the stellar tidal tails and galactic halos. 
Usually tails are rather prominent, may extend to radial distances above 100~kpc, before gradually dispersing themselves, becoming diffuse, or falling back on the host galaxy. 
Examples of prominent tidal tails are shown in Fig.~\ref{fig:tails}. 

\begin{figure}
\centerline{\includegraphics[width=0.6\columnwidth]{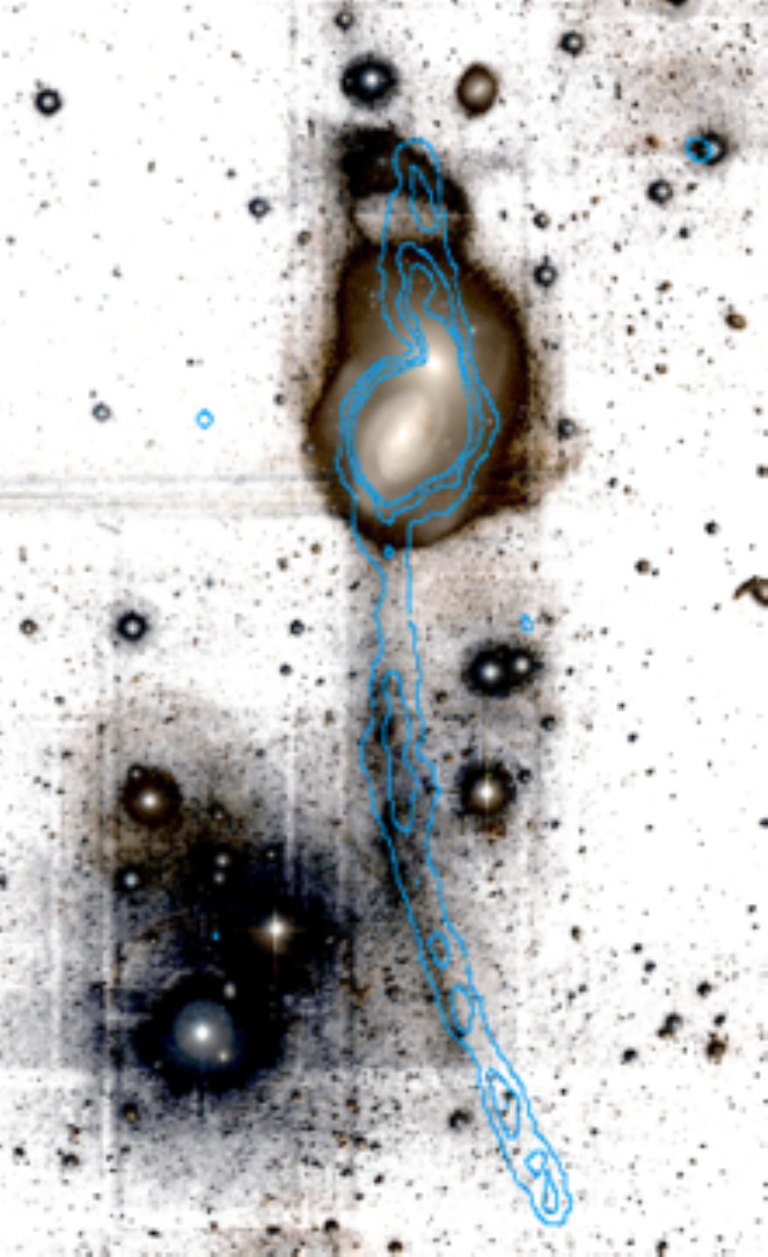}}
\caption{The \HI\ and stellar tidal tails of NGC~3226/27. \HI\ contours from the WSRT \citep{Serra12} are superimposed on a composite g--band plus true colour image of the system.  }
\label{fig:HI-tails}
\end{figure}

Such structures either trace on-going tidal interactions with a massive companion (the examples shown in Fig.~\ref{fig:tails}) or a past merger. 
Collisions between  pressure supported bodies are inefficient at forming tidal tails. Thus their detection implies that their progenitors had initially a  dynamically cold, and in most cases gas--rich  disc component. As a consequence, tidal tails are usually tracers of wet, major mergers. Gas clouds, mainly \HI\, but also molecular gas \citep{Braine01}, are expected to be associated with the stellar component of the tails, and were indeed detected in some stellar tails of ETGs \citep{Duc11}. In that case, such clouds may locally collapse, form stars or even (tidal) dwarf galaxies that will appear as  blue condensations within  tails of redder colour mainly composed of old stars \citep{Duc14}. 

However, the close physical connection between a stellar and \HI\ tail, although  frequent -- see the example of NGC~3226/27 in Fig.~\ref{fig:HI-tails} --, is not compulsory. Indeed, in addition to tidal forces that affect stars and gas the same way, the gaseous component may react to additional  environmental effects, such as ram pressure. As a result the \HI\ and stellar tails may be separated \citep{Mihos01}. Besides, depending on local conditions both components do not have the same life expectancy, and one may become invisible before the other.

\subsubsection{Tidal streams}

\begin{figure*}
\includegraphics[width=\textwidth]{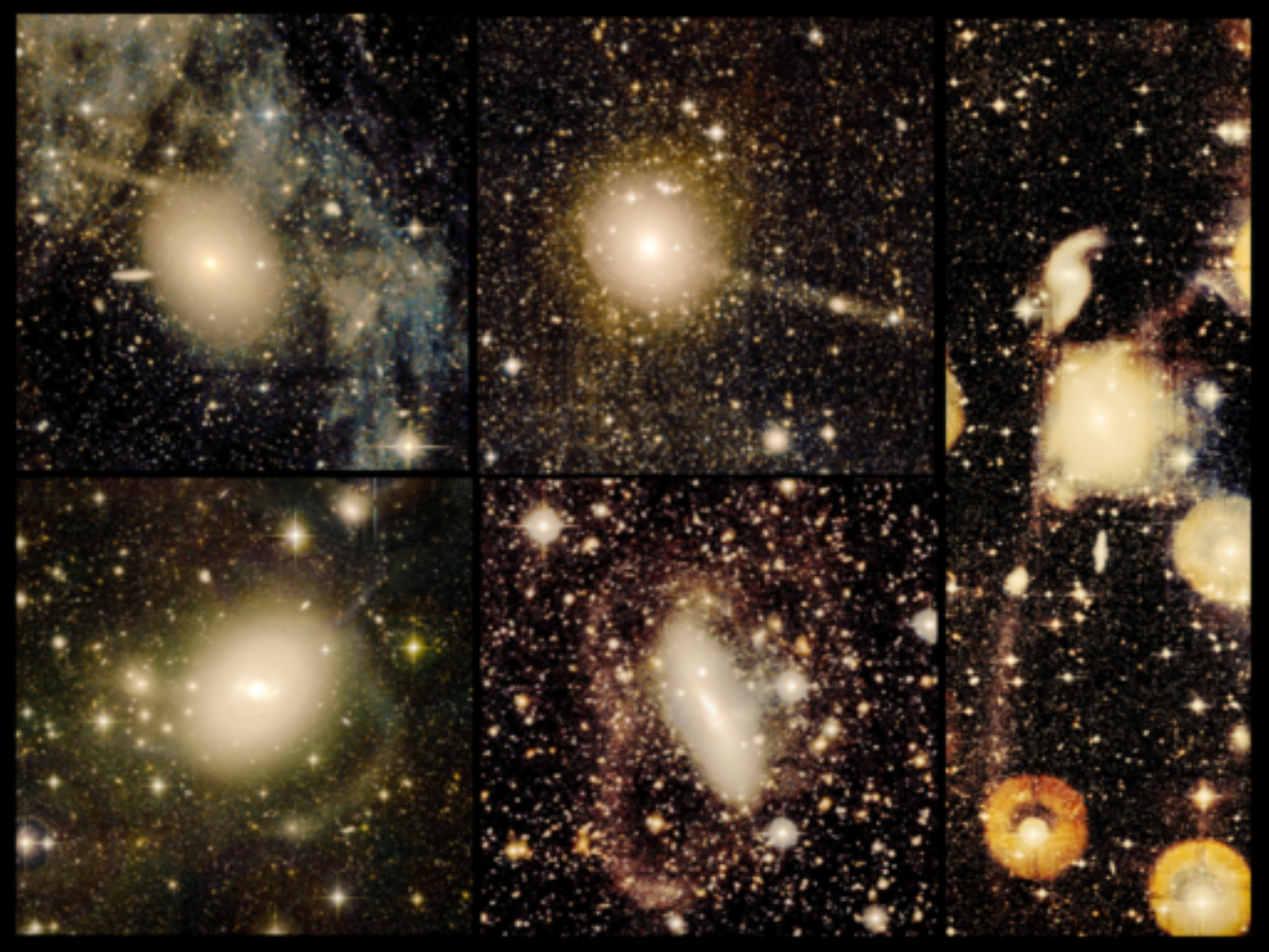}
\caption{Examples of ETGs hosting tidally disrupted satellites, as indicated by the presence of stellar streams around them. Clockwise, true colour images of NGC~2592, NGC~5198, NGC~3414, IC~1024 and NGC~936. Prominent cirrus emission is observed in the field of NGC~2592.}
\label{fig:streams}
\end{figure*}

We refer to tidal streams the structures that emanate from a low mass companion which is currently orbiting the primary galaxy or  is being ingested by it.  We thus adopt the point of view of the most massive galaxy with that definition. Contrary to the tidal tails, the material in the tidal streams differs from that of the main galaxy.  Having been expelled from the companion, their stars  have thus generally a lower metallicity. As a consequence, on the deep images, streams should be bluer than tails, though in practice the lack of signal in these low surface brightness features makes the comparison rather difficult.
Besides, streams appear as narrow and possibly very long filaments. Their length and shape are not only the result of tidal forces. They trace the orbit of the satellite around its massive host.
Examples of tidal streams are presented in Fig.~\ref{fig:streams}. In all of them, but IC~1024, a bright condensation is visible. It most likely corresponds to the remnant of the tidally disrupted companion. 
The linear shape of the streams around NGC~2592 and NGC~5198 suggests a radial recent collision, while the curved shape of the streams around NGC~936 and IC~1024 indicate that the companion has wrapped around its host for  a much longer time. 

With the adopted definition, streams trace past minor mergers, with a possible bias towards dry ones, given that the satellites of early-type galaxies tend to be   gas--poor, having lost their gas by internal -- feedback --  or external  -- ram pressure -- processes before they were accreted \citep[e.g.][]{Geha06}. Consequently, contrary to tidal tails, tidal streams are generally not expected to have any \HI\ counterpart and host star--forming regions.

\subsubsection{Shells}

Shells, also known as ripples, are identified by their circular shapes and shape-edged inner structure. Whenever present around a galaxy, they are usually numerous and for at least one class of shells, concentric. Despite an abundant literature devoted to shells and multiple models aimed at reproducing them, their origin is not yet fully understood. Indeed different types of collisions may produce them \citep[see the review by ][and references therein]{Struck99}.  It is relevant for our study to note that the physical conditions in intermediate mass mergers favour the formation of shells. 

\begin{figure*}
\includegraphics[width=\textwidth]{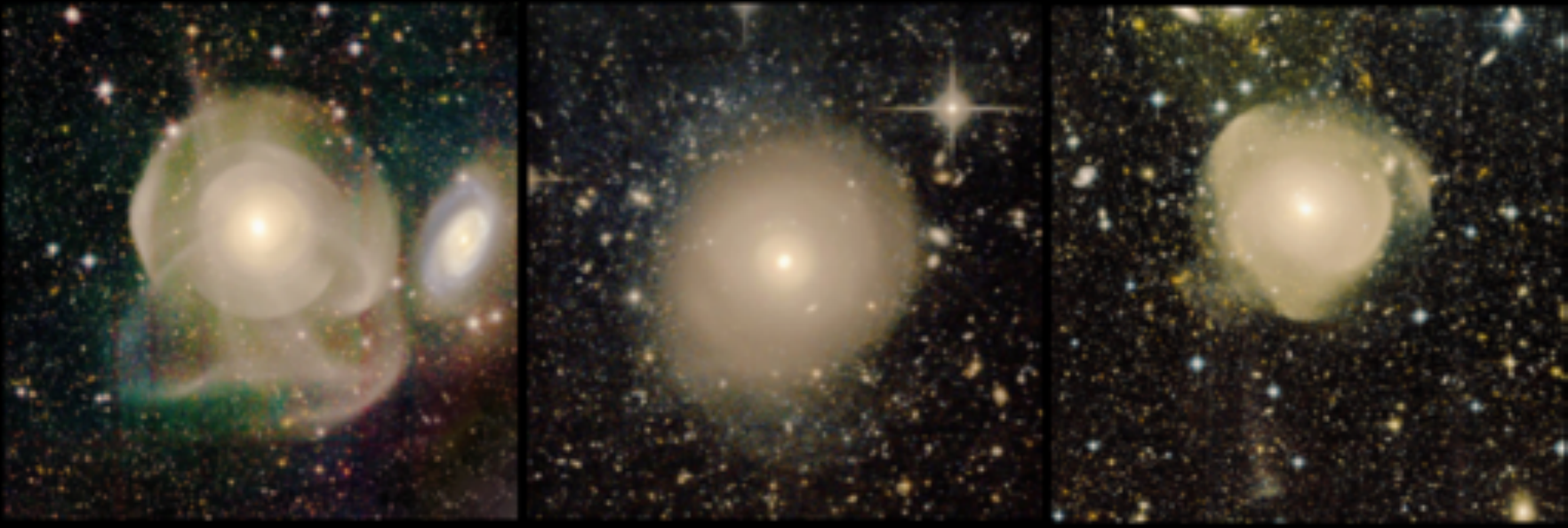}
\caption{Examples of  ETGs  exhibiting shells around them. From left to right, true colour images of  NGC~474,  NGC~502 and NGC~3619.}
\label{fig:shells}
\end{figure*}

Examples of galaxies in our sample showing prominent shells around them are presented in Fig.~\ref{fig:shells}. Several  of the galaxies in our sample, in particularly the truly spectacular galaxy NGC~474, were already known to exhibit shells \citep{Turnbull99}. Indeed, having not a particularly  low surface brightness, they were visible in previous generations of imaging surveys. Our deep imaging survey reveals new shells, located in the outermost regions, as well as so far unknown radial linear structures which may have formed at the same time. These structures might provide new constraints for numerical models. The MegaCam images also disclose a variety of colours from one shell to the other -- see in particular NGC~474 --, which should be taken into account when modeling the collision.

\begin{figure*}
\includegraphics[width=\textwidth]{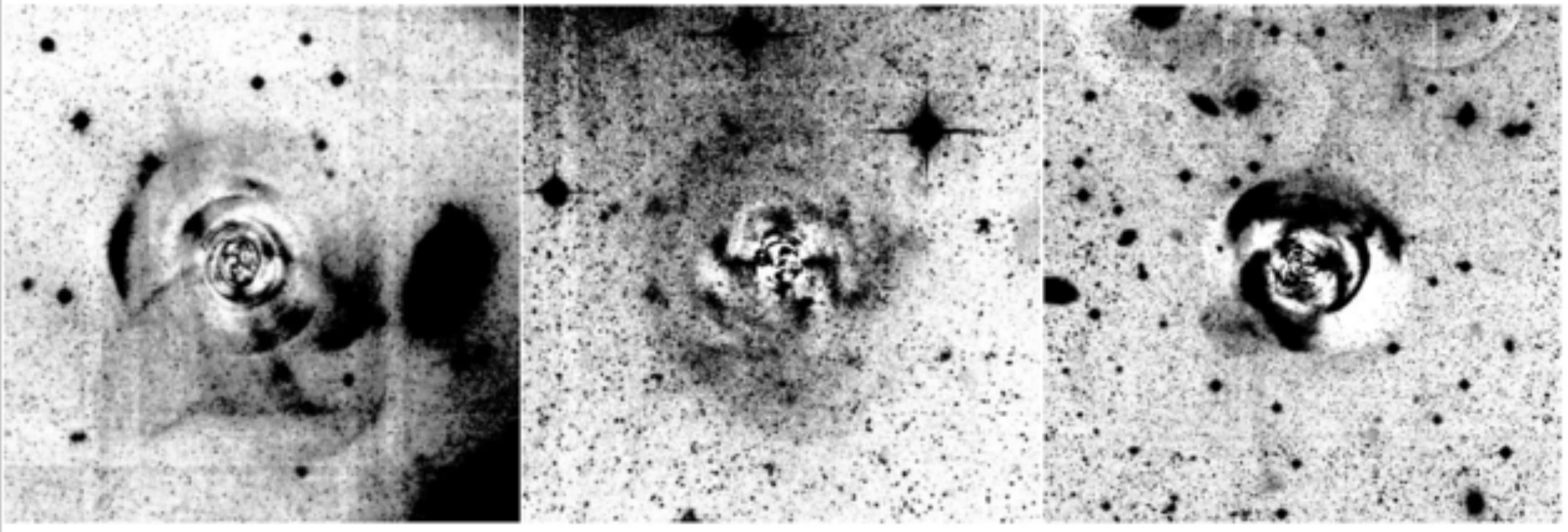}
\caption{Shells disclosed by subtracting a galaxy model with ellipse fitting. From left to right, residual g--band images of  NGC~474,  NGC~502 and NGC~3619. Note also the presence of radial structures,  formed either together with the shells  during the merger of the companion  (likely the case for NGC~474) or during a previous or late independent accretion event (NGC~3619). }
\label{fig:shells-sub}
\end{figure*}

Unsharp masking techniques have traditionally been used to disclose the  inner shells embedded in the body of early-type galaxies. We have rather used here an ellipse fitting modeling and subtraction (see Sect.~\ref{sec:model}) to reveal them. The technique  offers the advantage of enhancing the shells while preserving the more diffuse tidal tails and streams \citep{Forbes92}. 
Residual images of the shell galaxies presented in Fig.~\ref{fig:shells} are shown in Fig.~\ref{fig:shells-sub}. Besides the ripples, some of them also show radial streams.

Shell identification is usually unambiguous, especially because of their specific shape and multiplicity. Nonetheless in some cases, they may be confused with wrapping streams, or rings.

\subsection{Eye classification}
Automatic  algorithm  parametrizing galaxies with coefficients such as the  CAS \citep{Conselice03c}, Gini \citep{Abraham03}, or M20 \citep{Lotz04}, have been widely used for morphological classification of large number of galaxies, especially distant ones.
Whereas they may identify highly perturbed galaxies, such as the post-mergers presented in Sect.~\ref{sec:mergers}, they are not sensitive to the LSB components of galaxies, and thus are not able to distinguish the presence or absence of fine structures around apparently relaxed  systems. Furthermore, as argued above, identifying each  type of fine structure is particularly instructive and allows us, confronting the results with predictions of numerical simulations,   to  trace back the various types of mergers responsible for  the mass assembly  of galaxies. 

Machines would have a hard time dealing with  the subtlety of for instance disentangling a stream resulting from a minor merger from a tail, vestige of a major merger. In such conditions, eye classification remains unavoidable. 
  In a second step, the use of machine learning algorithms \cite[e.g.][]{Huertas-Company13} that are trained to reproduce the output of the eye classification,  may be considered.

 The Galaxy zoo project has pioneered the eye classification of large numbers of galaxies with equally large numbers of volunteers filling web--based polls  \citep{Lintott08,Willett13}. More recently the CANDELS project also used on--line  tools, but with a smaller number of  classifiers consisting of professional astronomers \citep{Kartaltepe14}. 
We used the latter approach to exploit our survey.
 Team members were invited to browse with on--line navigation tools a set of images similar to that presented in the catalog, consisting of true colour images, surface brightness, colour maps and residual images,  to identify and whenever possible count  the external  features.
On average, each galaxy was classified by 8 team members. A statistical analysis was  then performed, and galaxies were  assigned a  type characterizing  the properties of their outermost regions, i.e. shape of their  halos, frequency and type of fine structures. 

The first capital letter of the assigned type summarizes the global status of the galaxy, indicating whether it is fully regular ('R'), involved in an on-going tidal interaction with a massive companion ('I'), in an on-going or past collision with a low-mass companion (C), or show evidence for being a post-merger (M).    
Table~\ref{tab:code-status} lists the criteria used to assign the galaxy types and gives references to the figures illustrating them.  
In addition,  the presence of various types of features around the galaxies is coded:   a  '+' indicates the presence of   stellar structures like ripples (+r), tails (+t), shells (+s), discs (+d) or  perturbed halos (+h);   a  '-'  indicates contamination  by the halos of nearby bright objects (stars, companion galaxies, -h) or Galactic cirrus (-c). 
Further details are given in Table~\ref{tab:code-feature}. 
As  an example, a galaxy classified as I+t+ph-h  like NGC~5574 (see Fig.~\ref{fig:tails}) is involved in an on-going  interaction with a massive companion, exhibits tidal tails, a perturbed halo and is embedded within the halo of a bright object; a galaxy  classified as R-sc like NGC~509  (see Fig.~\ref{fig:sb}) is apparently fully relaxed but strong cirrus emission is present in its vicinity. 
In this classification scheme, an on-going interaction (I) takes precedence over a post-merger (M) which itself takes precedence over a minor merger (C). 
Thus a galaxy believed to be involved in an on-going interaction, like NGC~3414  (see Fig.~\ref{fig:streams}), but also showing streams from a disrupted companion, is classified as 'I' rather than 'C'.  This scale of priorities corresponds to a decreasing level of expected tidal perturbations on the stellar populations of the galaxy.

\begin{table}
\caption{Criteria used for the galaxy classification}
\begin{tabular}{llp{0.4\columnwidth}l}
\hline
Code  &   Type &  Description & Illustration \\ \hline
 R &  Fully relaxed  &  Regular  halo; no fine structure &   Fig.~\ref{fig:relaxed}\\
 C & Minor merger & Regular  halo; streams or shells from an accreted low-mass companion & Fig.~\ref{fig:streams} \\
 M & Major merger & Strongly perturbed halo; dust lanes; tidal tails; no massive companion &  Fig.~\ref{fig:mergers} \\
 I & Interacting & Perturbed halo; prominent tails due to a tidal interaction with a massive companion   &  Fig.~\ref{fig:tails} \\
U & Undetermined & Too close to a bright halo or Galactic cirrus to assign a type & \\  \hline
\end{tabular}
\label{tab:code-status}
\end{table}

\begin{table}
\caption{Coding of LSB structures and contaminants}
\begin{tabular}{lp{0.65\columnwidth}l}
\hline
Code  & Features / contaminants   & Illustration \\ \hline
+s &   stream &  Fig.~\ref{fig:streams} \\
+r &  shells / ripples & Fig.~\ref{fig:shells} \\
+t &   tail & Fig.~\ref{fig:tails} \\
+d & external star-forming disc / ring & Fig.~\ref{fig:disk}  \\
+ah & asymmetric halo & \\
+ph & perturbed halo & \\
+wl & weak central dust lanes & \\
+pl & prominent dust lanes  & \\ \hline
-h & galaxy embedded in the halo of a nearby star or galaxy & Fig.~\ref{fig:halos} \\
-wc & weak Galactic cirrus in the field & \\
-pc & prominent  Galactic cirrus & Fig.~\ref{fig:cirrus} \\ \hline
? & presence of a given feature is uncertain & \\ \hline
\end{tabular}
\label{tab:code-feature}
\end{table}

Table~\ref{tab:class}  lists the adopted classification and provides individual comments for the first galaxies in the catalog. The full table is available in the online web version.  
Rough statistics on the classification -- number of galaxies and percentage -- is provided by Table~\ref{tab:stat}.  
In  the sample presented here, half of the ETGs show some sort of tidal perturbation whereas  35~\%\ appear to be  regular, even at the depth of the survey. These fully relaxed galaxies are highlighted in green  on the diagrams stellar mass vs volume density (Fig.~\ref{fig:loc}),  dynamical mass vs effective radius (Fig.~\ref{fig:MRe}) and specific angular momentum vs ellipticity (Fig.~\ref{fig:rot}).
A tendency emerges from these figures:   the least massive and fast rotating galaxies seem to be  less perturbed than the  massive, slow rotating galaxies. A detailed analysis of the trends is postponed to other papers and the completion of the survey. Despite the large number of galaxies presented here with respect to other deep imaging surveys (92), they only corresponds to 35~\% of the \AD\ volume limited sample, and the catalog has selection biases that need to be taken into account in the interpretation of the data (see Sect.~\ref{sec:sample}).

\begin{table*}
\caption{ETG classification based on the deep imaging (first entries; full table available in the online web version)}
\begin{tabular}{llp{0.7\textwidth}}
\hline 
Galaxy & Class & Individual comments \\ 
\hline 
NGC0448 & I+s & The ETG is in a tidal interaction with a disturbed companion. \\ 
NGC0474 & M+s+r+ph & The ETG is surrounded by multiple concentric shells and hosts several radial streams. Its outer halo reaches the disk of the unperturbed companion spiral galaxy, NGC 0470. \\ 
NGC0502 & M+t?+r?+ah-wc-h & The stellar halo of the ETG is asymmetric, possibly due to the presence of a diffuse tidal tail and/or a shell. \\ 
\hline

\end{tabular} 
\label{tab:class}
\end{table*}

\begin{table}
\caption{Statistics on galaxy classification}
\begin{tabular}{lcc}
\hline
Class / features & Number of ETGs & Fraction \\ \hline
Relaxed (R) & 32 & 35~\%  \\
Minor merger (C) & 15 & 16~\%  \\
Major merger (M) & 11 & 12~\%  \\
Interacting (I) & 20 & 22~\%  \\
Undetermined (U) & 14 & 15~\%  \\ \hline
with stream(s) & 26 & 28~\%  \\
with tail(s) & 16 & 17~\% \\
with shell(s) & 19 & 21~\% \\ 
with a perturbed/asymmetric halo & 32 & 35~\% \\
with a star forming disc & 6 &  7~\% \\
\hline
\end{tabular}
\label{tab:stat}
\end{table}

\subsection{The structural parameters revisited}

 \subsubsection{The contribution of the outer halo to the total luminosity}
 
 Some of the extended structures around ETGs revealed by the MegaCam images look rather prominent on surface brightness maps (see Fig.~\ref{fig:sb}).  In fact,  they are  of low luminosity and  do not contribute much to the total mass of the galaxy.
We computed from the model galaxies (obtained with the ellipse fitting) the relative faction of g--band flux enclosed  by the isophote   26~\sbr and that corresponding to the  minimum isophote level below which  the ellipse fitting algorithm failed (on average  27.7~\sbr) \footnote{This is not the limiting surface surface brightness, which is deeper. Besides galaxies for which the ellipse fitting failed already below 27~\sbr were excluded from the analysis: they correspond to galaxies contaminated by the halos of bright stars which could not be properly subtracted.}.
We found that this  fraction of extra g--band light revealed by our survey corresponds on average to only 5~\% of the total light of the galaxy. The maximum is  16~\%. In the r--band, the excess light, below the isophote 25.5~\sbr  (to take into account the color term)  is similar: 4.5~\%, with a maximum of 14~\%.   
This value does not include the fine structures -- streams, tails or shells --, that are de facto excluded by the ellipse fitting  algorithm. Their contribution may be determined from aperture photometry, albeit with very large error bars given their very low surface brightness. 

The rather  low contribution of the outer halo to the total luminosity   implies that the structural parameters of the ETGs in our sample -- mass, effective radius -- initially derived from the shallow SDSS and INT images should not change dramatically when measured on the MegaCam images.

\subsubsection{Changes in effective radii from deep imaging}

\begin{figure}
\includegraphics[angle=-90,width=\columnwidth]{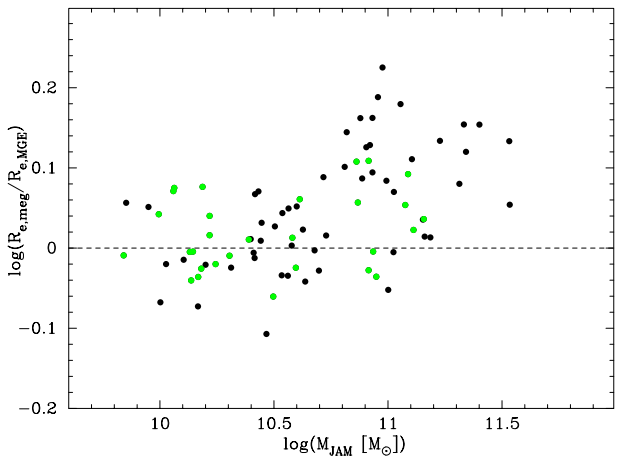}
\caption{Ratio between the effective radius estimated with the MegaCam r band (with an ellipse fitting modeling of the ETG) and that determined from the SDSS/INT images  \citep[with  MGE modeling,][]{Cappellari13} as a function of the dynamical mass, M$_{\rm JAM}$. Relaxed galaxies showing no sign of tidal perturbations even with the deep imaging are shown in green.}
\label{fig:Re}
\end{figure}

We compared  the  effective radius directly determined from the ellipse model of the ETG, R$_{\rm e,meg}$\footnote{based on the MegaCam r--band image, along the major axis. The total luminosity was obtained within the last isophote fitted by the ellipse procedure.} with the effective radius computed from the MGE models by \cite{Scott13}, R$_{\rm e,MGE}$\footnote{tabulated in Table 1, column 10 of 
\cite{Cappellari13}. Like for the MegaCam analysis, these values are derived from the observed images, without extrapolating the photometry to infinite radii. For a consistent comparison, they were not multiplied by the factor of 1.35 used to reconcile the MGE value with the original one in \cite{Cappellari11}.}. On average, the MegaCam  values are just 11~\% larger. As seen in Fig.~\ref{fig:Re},  below  $5 \x 10^{10}~\Mo$, the old and revised measures of \Reff\ agree, with a scatter of about 0.1, i.e. within the typically 20\% measurement error of  \Reff  \citep{Cappellari13}.
The galaxies with unchanged measures are those we classified as ``relaxed (R)''  in our morphological analysis. 
Above  a transition mass of $\sim 10^{11}~\Mo$, the MegaCam values of   \Reff\  are systematically higher, with a maximum excess of a factor of 1.7. 
For these galaxies, the   larger fraction of stellar light below the SDSS surface brightness limit detected by MegaCam accounts for the  increase of \Reff.
This is also related to the observation that massive galaxies are best fitted by larger \cite{Sersic68} indices \citep[e.g.][]{Caon93,Kormendy09}.
Note that similar transition masses are also observed for other ETG properties such as the internal kinematics 
\citep[see Fig.~14 in][]{Cappellari13b}.

\section{Summary and perspectives}
\label{sec:sum}

We have presented an image  atlas of  92 early-type galaxies taken from the \AD\ volume limited sample. The ETGs  have been observed with the MegaCam camera on the Canada-France-Hawaii-Telescope. The observing strategy and data reduction were optimized for the detection of extended low surface brightness (LSB) structures. 
The resulting images have a limiting surface brightness in the g--band at least 2 mag fainter than  so far available for this sample. 
The catalog (available in electronic format) includes ETGs located in a low to medium density environment (thus excluding the Virgo Cluster) and spans the whole range of masses and kinematical properties probed by the \AD\ survey.
 g and r band images are available for all galaxies in this catalog, while a fraction of them have also been observed in the u and i band.
 The atlas consists of true colour composite images, surface brightness and colour maps, residual images resulting from the subtraction of a model of the ETG obtained with ellipse fitting. An image of the large scale  environment  of each galaxy  covering one square degree is also shown.  
 
These images reveal a number of LSB structures around the  ETGs which were not visible on the previous generation of images, including the SDSS.
We discuss how deep imaging may  change or not the way we see and classify these galaxies. In particular the presence of blue star--forming spiral structures around several objects so far considered as red and dead may appear troublesome. It reminds us about the fragility of a morphological classification   based on inspection of images having different depth and of the need to add other criteria, such as the internal kinematics as proposed by the  Sauron and \AD\  collaborations.\\

The previously unknown fine structures were arranged in different classes:\\ 
$\bullet$  prominent tidal tails, that may be  gas--rich, and  share the same colour as their host apart from local condensations where in-situ star-formation proceeded. They trace  on going tidal interactions with massive companions or past  major mergers.\\
$\bullet$  long and narrow tidal streams, sometimes wrapping around the ETG. The progenitor of their stars is likely a gas--poor low--mass disrupted companion. Its remnant is often still visible somewhere along the filament. These streams  made of low metallicity material  are expected to have a different colour than the host. They  trace minor  mergers.\\
$\bullet$ shells or  ripples, that surround the ETG, often as a series of concentric circles. Such sharp-edge structures, which are best disclosed subtracting the host, were usually already visible on shallow images. However our survey revealed new ones at larger radii  and  associated  radial linear structures. At least a fraction of them trace intermediate mass mergers.

The census of each type of fine structures was carried out by a visual inspection made by the team members. We discuss the  ambiguities of such identification and present the prospects of automatic identification and classification. 

We  note  that a rather large fraction of early-type galaxies -- about one third in our sample -- remains fully regular even at the depth of our imaging survey. Whether, given  our sensitivity limit, their lack of external LSB structures  and vestiges of past collisions  is compatible with predictions from the  cosmological simulations should be further investigated.  
For most ETGs, the extra diffuse component disclosed by the survey does not increase  significantly  the total stellar budget. On average, isophotes fainter than 26~\sbr in the g--band  contribute to about 5 percent, and a maximum of 16 percent of the total luminosity of the galaxy, not taking into account the external fine structures. An increase by on average 11\% of the effective radius of the galaxies is obtained with MegaCam with respect to the previously published values.  Above a transition mass of $\sim 10^{11}~\Mo$, the excess is systematic, with a maximum  factor of 1.7.

The detailed statistical  analysis of the results  will be presented in future papers. They will address  correlations between the degree of tidal perturbation with both the large scale environment and  the   properties of ETGs (internal kinematics and structure, gas content, etc...), which are already known thanks to the wealth of multi-wavelength data collected as part of the \AD\ project. Besides,  the follow-up and on-going CFHT Large Programme MATLAS, which has exactly  the same observing strategy as the one presented here,   will  complete the deep imaging survey.  Multi-band images of hopefully all 260 \AD\ galaxies will be acquired. The  galaxies located in the Virgo cluster -- about one fourth of the \AD\ sample --  were already observed as part of the Next Generation Virgo Cluster Survey (NGVS) and an image atlas similar to the one presented here for the field ETGs will be presented in the NGVS paper series.

Several teams, some including amateur astronomers, are currently carrying out similar deep, LSB optimized,  imaging surveys. Ours does not necessarily stand out  by the depth reached:  about 28.5~\sbr in the g--band. Projects  like that announced by  \cite{vanDokkum14} claim significantly  deeper  limiting surface brightness, though as argued in this paper, determining its real value is not straightforward, as it is set by the  background variations rather than by photon noise statistics. Observing with a medium size telescope, rather than with a small telescope as usually   done in deep imaging experiments,  allowed us to limit the integration time to less than one hour per band, instead of a full night  or more, and thus to target a large number of galaxies.  Besides, the great Image Quality of MegaCam and the observations from a site exceptional for its low seeing -- the Mauna Kea --, enables to address additional scientific questions, such as the detection of globular clusters and determination of their distribution around the ETGs. This is another archeological probe of the assembly of galaxies \citep{Zhu14,Brodie14}.

Nonetheless,  a survey made with a  complex camera like MegaCam faces a number of difficulties, some being intrinsic to any LSB study. We have presented a number of them, and on-going efforts to minimize them. In particular internal reflections within the camera imprint on the CCDs extended ghost halos which badly affect the colour of galaxies.  Cumulated, they may contribute to a large fraction of the background and being the limiting factor for LSB science from the ground. 
Besides the exploration of the LSB extragalactic Universe is hampered by the presence of extended emission from Galactic cirrus. At the depth of the survey, the  scattered light from dust clouds becomes prominent, even at relatively large Galactic latitude. Their filamentary structure and colour are similar to that of tidal tails and streams, and thus cannot be easily subtracted. On the positive side,  this scattered light  offers the  opportunity to study the Milky Way cold dust distribution at unprecedented high  spatial resolution. 

Instrumental artefacts and cirrus contamination should however not prevent the exploitation of diffuse light for  galactic archeology, which has so far been underused. Even in the area of extremely large telescopes, the number of galaxies for which resolved stellar populations is accessible will remain limited. Beyond 10-20 Mpc, the LSB component of galaxies might be one of the most promising tracer of the mass assembly of galaxies, and its study provides key checks for numerical cosmological simulations.

\bibliographystyle{mn2e}
\bibliography{../Biblio/all}

\section*{ACKNOWLEDGEMENTS} 
The paper is based on observations obtained with MegaPrime/MegaCam, a joint project of CFHT and CEA/DAPNIA, at the Canada-France-Hawaii Telescope (CFHT), which is operated by the National Research Council (NRC) of Canada, the Institute National des Sciences de l'Univers of the Centre National de la Recherche Scientifique of France, and the University of Hawaii. 
All observations were made as part of the service  mode offered by the CFHT. We are grateful to the queue team for their dedication and for the anonymous referee. 
The project greatly benefited from the experience of the NGVS team members, in particular Laura Ferrarese, Patrick C\^{o}t\'e, Chris Mihos, and discussions  with Etienne Ferri\`ere,  Rodrigo Ibata and David Valls-Gabaud who contributed very much to the advance of LSB science. We are grateful to  Velimir Popov and Emil Ivanov who run the robotic amateur Irida observatory  in Bulgaria,  obtained and reduced the image of NGC~474 shown in Fig.~\ref{fig:pro-amateur}.
We warmly thank Herv\'e Dole who extracted  the Planck images we used to estimate cirrus contamination. 
This research has made use of  the NASA/IPAC Extragalactic Database (NED) which is operated  by the Jet Propulsion Laboratory, California Institute of Technology, under contract with the National Aeronautics and Space Administration.
PAD acknowledges support from Agence Nationale de la Recherche (ANR10-BLANC-0506-01). 
LMD acknowledges support from the Lyon Institute of Origins under grant ANR-10-LABX-66.
  MC acknowledges support from a Royal Society University Research Fellowship.
This work was supported by the rolling grants PP/E001114/1 and ST/H002456/1 and visitors grants PPA/V/S/2002/00553, PP/E001564/1 and ST/H504862/1 from the UK Research Councils. RLD acknowledges travel and computer grants from Christ Church, Oxford and support from the Royal Society in the form of a Wolfson Merit Award 502011.K502/jd. SK acknowledges support from the Royal Society Joint Projects Grant JP0869822.
TN  acknowledges support from the DFG Cluster of Excellence `Origin and Structure of the Universe'.
MS acknowledges support from a STFC Advanced Fellowship ST/F009186/1.
LY acknowledges support from NSF grant AST-1109803.
TAD acknowledges the support provided by an ESO fellowship. The research leading to these results has received funding from the European Community's Seventh Framework Programme (/FP7/2007-2013/) under grant agreement No 229517.
The authors acknowledge financial support from ESO.

\end{document}